\documentclass[submitting]{nst}

\usepackage{subfigure,dcolumn}
\usepackage[T2A,T1]{fontenc}
\usepackage[russian,english]{babel}
\usepackage{amsmath}
\usepackage{verbatim}
\usepackage{listings}
\usepackage{makecell}
\usepackage{mathrsfs}
\usepackage{amsbsy}

\usepackage{color}

\graphicspath{{HRplot/}}
\graphicspath{{detsim},{calib},{gaus},{HRplot},{figure}}

\begin{document}
\bibliographystyle{plain}

\title{Sub-GeV events energy reconstruction with 3-inch PMTs in JUNO}\thanks{ We thank the JUNO reconstruction and simulation working group for many helpful discussions. This work was supported by National Key R\&D Program of China No. 2023YFA1606103, National Natural Science Foundation of China No. 12375105 and 12005044, and the Guangxi Science and Technology Program (No. GuiKeAD21220037).}

\author{Si-yuan Zhang}
\affiliation{School of Physical Science and Technology, Guangxi University, Nanning 530004, China.}
\author{Yong-Bo Huang}
\email[Yong-Bo Huang, ]{huangyb@gxu.edu.cn}
\affiliation{School of Physical Science and Technology, Guangxi University, Nanning 530004, China.}
\author{Miao He}
\email[Miao He, ]{hem@ihep.ac.cn}
\affiliation{Institute of High Energy Physics, Beijing 100049, China.}
\author{Cheng-Feng Yang}
\affiliation{School of Physics, Sun Yat-Sen University, Guangzhou 510275, China.}
\author{Guo-ming Chen}
\affiliation{School of Physical Science and Technology, Guangxi University, Nanning 530004, China.}

\begin{abstract}
A 20-kiloton liquid scintillator detector is designed in the Jiangmen Underground Neutrino Observatory (JUNO) for multiple physics purposes, including the determination of the neutrino mass ordering through reactor neutrinos, as well as measuring supernova neutrinos, solar neutrinos, and atmosphere neutrinos to explore different physics topics. Efficient reconstruction algorithms are needed to achieve these physics goals in a wide energy range from MeV to GeV. In this paper, we present a novel method for reconstructing the energy of events using hit information from 3-inch photomultiplier tubes (PMTs) and the OCCUPANCY method. Our algorithm exhibits good performance in accurate energy reconstruction, validated with electron Monte Carlo samples spanning kinetic energies from 10~MeV to 1~GeV.

\end{abstract}


\maketitle

\section{Introduction}
\label{section:Introduction} 

    Liquid scintillator (LS) detectors are extensively used in nuclear physics and particle physics. Over the past few decades, LS detectors have played crucial roles in achieving remarkable scientific results in neutrino experiments~\cite{DayaBay-2012,DoubleChooz-2012,RENO-2012,KamLAND-2013,Borexino-2014}. The central detector (CD) of Jiangmen Underground Neutrino Observatory (JUNO) will be the world's largest liquid scintillator detector, aiming to probe multiple physics goals including determining the mass ordering of neutrinos and accurately measuring neutrino oscillation parameters through reactor antineutrinos, as well as observing supernova neutrinos, solar neutrinos, atmospheric neutrinos, etc~\cite{JUNO-PPNP}. JUNO has developed highly transparent LS and highly efficient PMTs with 78\% photo coverage. These optimizations are necessary for JUNO to achieve the key requirements for determining the neutrino mass ordering, including an unprecedented energy resolution of $3\%/\sqrt{E(\rm{MeV})}$ and better than 1\% energy scale uncertainty. Additionally, JUNO has developed a dual calorimetry technique~\cite{He-2017dhc,Jollet-2020fck} that can not only calibrate the non-linearity of the charge response of 20-inch PMTs, but also to enable the detector to operate over a larger dynamic energy range.
    
    Meanwhile, efficient algorithms are necessary for the reconstruction of individual event energies in JUNO. For the energy region of the inverse beta decay (IBD) of reactor neutrinos ($E_{\rm{vis}}$ < 10~MeV), JUNO has developed many robust reconstruction algorithms based on traditional methods~\cite{Wu-2018zwk,Huang-2021baf,Huang-2022zum} or the machine learning technology~\cite{Qian-2021vnh,Gavrikov-2021ktt}. Events with visible energies larger than several hundreds of MeV in the detector are primarily from atmospheric neutrino interactions, and their energy and direction reconstructions have been well studied using both the probabilistic unfolding method~\cite{Settanta-2019ecp,JUNO-2021tll} and the machine learning technology~\cite{Wirth-2022mgi,ColomerMolla-2023ota} to assist in determining the neutrino mass ordering~\cite{juno-yellowbook}. However, energy reconstruction in the mid-energy region (10~MeV < $E_{\rm{vis}}$ < several hundreds of MeV), which includes events from diffuse supernova neutrino background (DSNB)~\cite{JUNO-2022lpc}, Michel electrons, low-energy atmospheric neutrinos, and possible proton decays~\cite{JUNO-2022qgr}, is rarely discussed in previous studies.
    
    The basic idea of current energy reconstruction algorithms in the JUNO central detector are based on either a data-driven maximum likelihood method or a machine learning strategy that utilizes the information from the detector hit pattern. The data-driven maximum likelihood method has the advantage of better modeling the response of the real detector. It primarily utilizes the detector response to radioactive sources as calibration templates, with the energy of radioactive sources being in the MeV range~\cite{Huang-2022zum}. However, it is found that the accurate energy reconstruction below 10~MeV can be affected by the spatial scale of energy deposition. Therefore, it is necessary to investigate the feasibility and performance when applying this method to cases where events cannot be treated as point-like. We introduce the second moment $S$ (Eq.~\ref{eq:second_moment}) to describe the shape of both point-like and cluster-like events. This physical quantity is commonly used in accelerator experiments to describe the shape of clusters in energy calorimetry.

    \begin{equation}
    \label{eq:second_moment}
    \begin{aligned}
        S=\frac{\sum_{\alpha = 1}^{N_{E}}{E_{\alpha} \times [\overrightarrow{r_{\alpha}}(x_{\alpha},y_{\alpha},z_{\alpha})} - \overrightarrow{r}(x,y,z)]^2}{\sum_{\alpha=1}^{N_{E}}{E_{\alpha}}}
    \end{aligned}
    \end{equation}

    where $N_{E}$ is the number of secondary energy depositions for the event. $E_{\alpha}$ and $\overrightarrow{r_{\alpha}}(x_{\alpha},y_{\alpha},z_{\alpha})$ are the energy deposition and position in the $\alpha^{\rm th}$ secondary energy deposition, respectively. $\overrightarrow{r}(x,y,z)$ is the energy-deposit center for the event, which is the weighted average of secondary energy deposition and can be calculated as follows.

    \begin{equation}
    \label{eq:edepCenter}
    \begin{aligned}
        \overrightarrow{r}(x,y,z)=\frac{\sum_{\alpha = 1}^{N_{E}}{\overrightarrow{r_{\alpha}}(x_{\alpha},y_{\alpha},z_{\alpha}) \times E_{\alpha}}}{\sum_{\alpha=1}^{N_{E}}{E_{\alpha}}}
    \end{aligned}
    \end{equation}
 
\begin{figure}[!htb]
        \centering  \includegraphics[width=0.9\hsize]{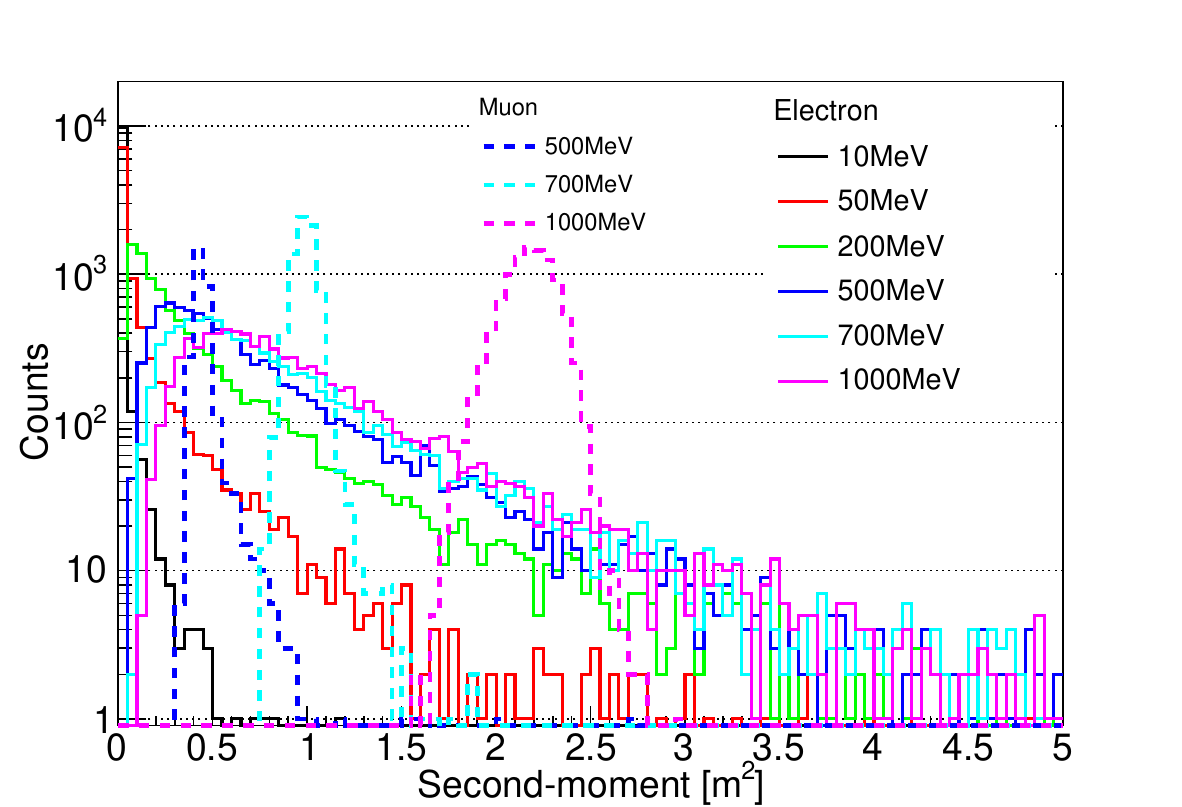}
        \caption{Distributions of the second-moment for electrons with different kinetic energies.}
        \label{fig:second_moment}
    \end{figure}

    Figure~\ref{fig:second_moment} compares the distributions of the second moment for electrons with different kinetic energies in the JUNO's LS. As the kinetic energy of the electron increases, the distribution of the second moment becomes more diffuse, which indicates that it corresponds to a larger cluster of energy deposition. For comparison, Fig.~\ref{fig:second_moment} also shows the second-moment distribution of muons with different energies deposited in the LS. It can be found that the second moment distribution of the cluster-like events and the track-like events with the same energy deposition is very different, which is related to their shape differences in energy deposition. On the other hand, for energy reconstruction of high-energy events, the potential deviation of PMT's reconstructed charge (charge non-linearity) is also an important issue to be considered.

    The machine learning approach demonstrates good performance in event reconstruction as it can effectively utilize the detector response information~\cite{Qian-2021vnh,Gavrikov-2021ktt,Wirth-2022mgi,ColomerMolla-2023ota}. However, it is important to note that this method relies on pure training sample from data or simulation. The former requires the reliable selection strategy and data accumulation, while the latter usually needs adjustments according to real data, especially in the early stages of detector operation. Similar to the data-driven maximum likelihood method, since the PMT hit pattern is the basic input of machine learning, investigating the influence of potential charge non-linearity of PMTs based on real data is necessary. 
    
    In this paper, we investigate the energy reconstruction of events spanning from MeV to GeV in the JUNO CD using the data-driven maximum likelihood method~\cite{Huang-2022zum}. To reduce the dependence on the accuracy and non-linearity of PMT charge reconstruction, we only utilize information of PMT firing states (fired or unfired), known as the OCCUPANCY method. This paper focuses on the reconstruction of cluster-like events since the calibration templates in our study are constructed using point-like (or cluster-like) calibration sources. Thanks to the study of event identification in JUNO~\cite{Settanta-2019ecp}, we can effectively distinguish cluster-like events from track-like events. As for the reconstruction of track-like events (mainly muon-like events that have long tracks in the detector), it is also an important topic and has been carried out in~\cite{Settanta-2019ecp,JUNO-2021tll,Wirth-2022mgi,ColomerMolla-2023ota,Genster-2018caz,Zhang-2018kag,Liu-2021okf,Yang-2022din}; however, it is not the subject of this article. To investigate the performance of our reconstruction, Monte Carlo (MC) simulation data generated by the JUNO offline software~\cite{Li_2017zku,Li-2018fny,Lin:2022htc} is used for validation. 
    
    The details of our study are presented as follows: First, we introduce the JUNO detector and 3-inch PMT system (Sec.~\ref{section:Detector}). Then, we present the methodology of our reconstruction (Sec.~\ref{section:Method}), including the construction of calibration maps and the construction of maximums likelihood function. In Sec.~\ref{section:Result}, the reconstruction performance will be shown and compared. Finally, a summary will be provided in Sec.~\ref{section:Summary}.

\section{JUNO detector and 3-inch PMT system}
\label{section:Detector}
\begin{figure}[!htb]
        \centering
        \includegraphics[width=1.2\linewidth]{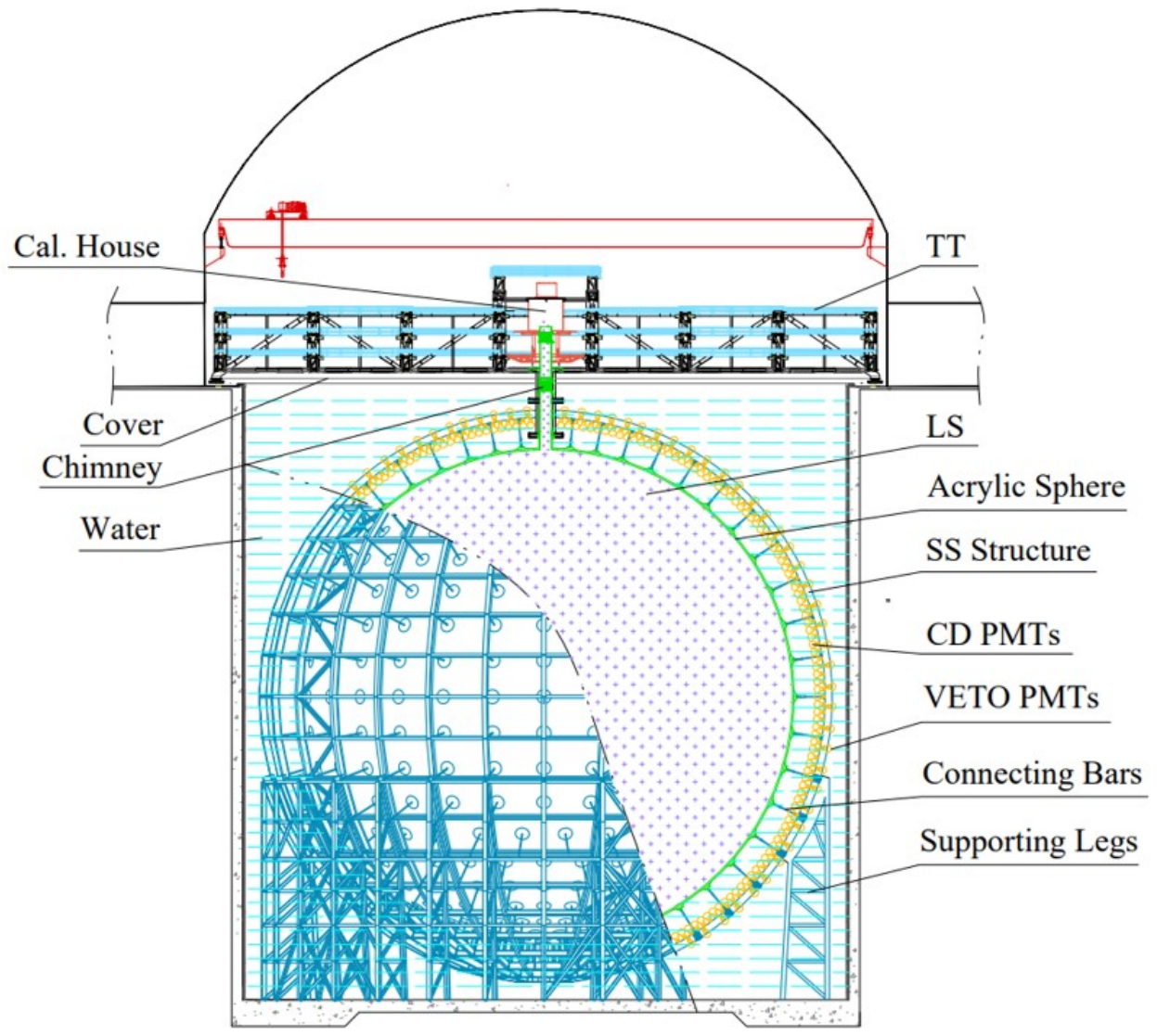}
        \caption{A schematic view of the JUNO detector~\cite{JUNO-PPNP}.}
        \label{fig:JUNO_schematic}
    \end{figure}

    As shown in Fig.~\ref{fig:JUNO_schematic}, JUNO mainly consists of three sub-detectors: CD, the water Cherenkov detector, and the top tracker detector~\cite{JUNO-2015sjr,JUNO-PPNP}. CD contains 20 kilotons of LS and a 12~cm thick acrylic spherical container with a diameter of 35.4~m. The main component of the LS is linear alkyl benzene(LAB), with PPO (2.5-diphenyloxazole) as fluor and bis-MSB as wavelength shifter. A total of 17612 20-inch PMTs (LPMTs) and 25600 3-inch PMTs (SPMTs) will be installed on the exterior of the container as photosensors to collect photon signals. As a result, more than 1345 photoelectrons (PEs) will be observed by CD for a 1~MeV electron that fully deposits its kinetic energy in the LS. The SPMTs will work almost exclusively in the single photoelectron (spe) mode for reactor antineutrino detection ($E_{\rm{vis}}$ < 10~MeV). Therefore, SPMTs can serve as a linear reference for LPMTs and be used to calibrate the charge non-linearity of LPMTs. This feature is helpful in constraining some of the systematic uncertainties in the LPMT energy reconstruction and improving the energy resolution. Moreover, SPMTs have the potential to detect supernova neutrinos and measure the solar parameters ($\theta_{12}$ and $\Delta m^{2}_{21}$) independently~\cite{JUNO:2022mxj}.

    \begin{figure*}[!htb]
    \centering  
    \subfigure[nPE received by LPMTs for 500~MeV electrons]{
    \label{fig:nPEbyPMT_500MeV_LPMTs}
    \includegraphics[width=0.3\hsize]{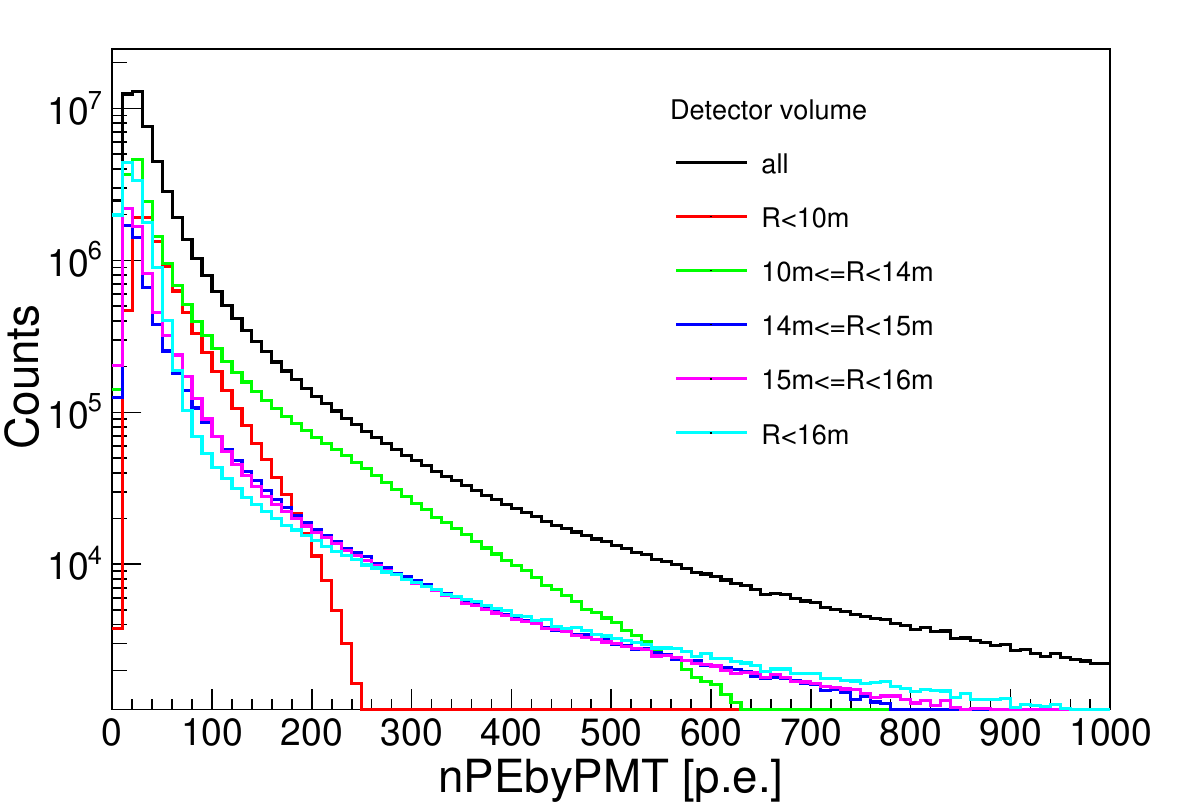}}
    \quad
    \subfigure[nPE received by SPMTs for 500~MeV electrons]{
    \label{fig:nPEbyPMT_500MeV_SPMTs}
    \includegraphics[width=0.3\hsize]{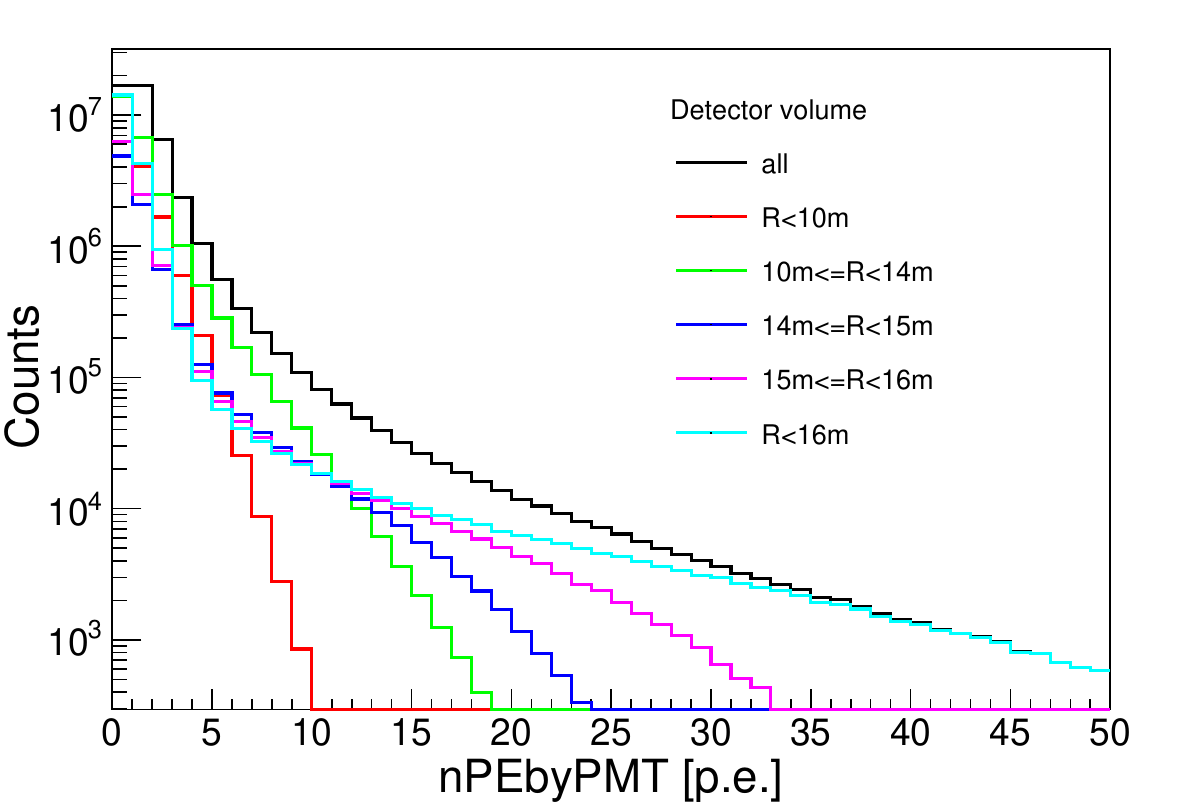}}
    \quad
    \subfigure[The proportion of fired SPMTs]{
    \label{fig:fireRatio}
    \includegraphics[width=0.3\hsize]{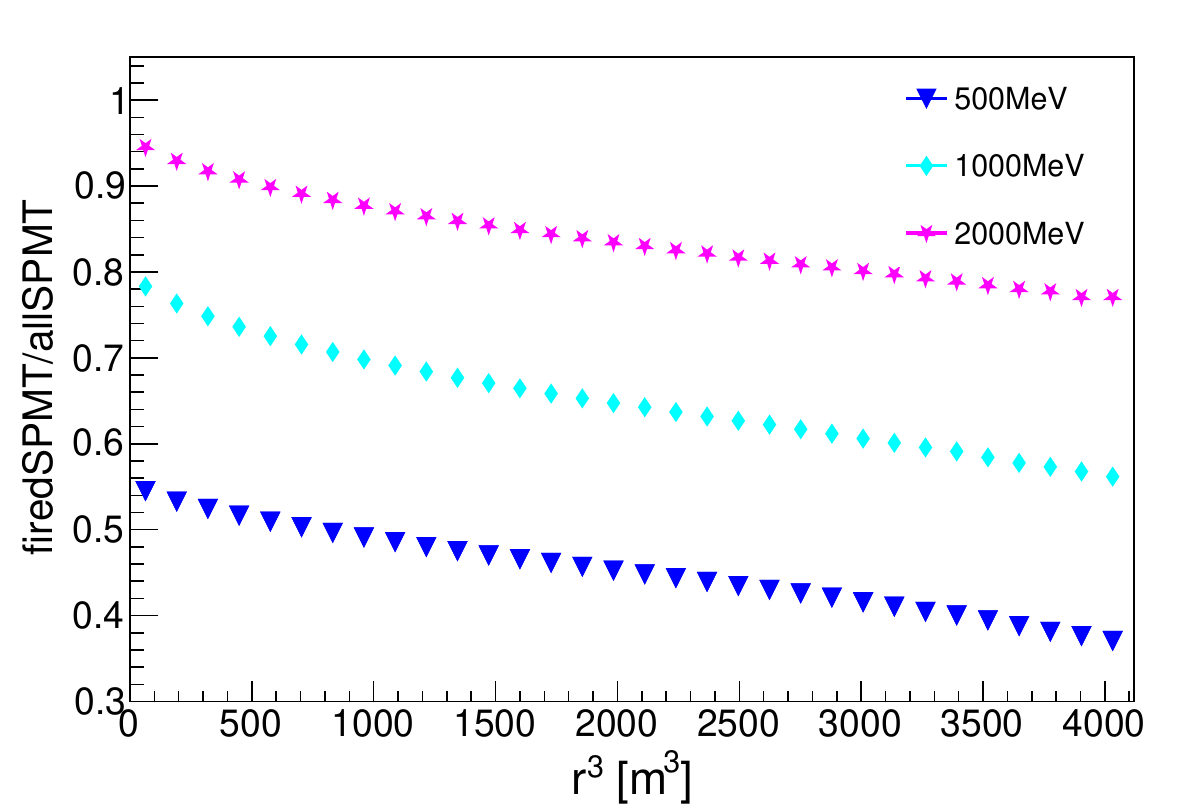}
    }
    \caption{The distribution of nPE received by LPMTs (a) and SPMTs (b) for 500~MeV electrons deposited their kinetic energies in the LS. (c) The proportion of fired SPMTs for electrons deposited their kinetic energies at different locations.}
    \label{fig:nPEbyPMT_500MeV}
    \end{figure*}
    
    For the detection of events with energies greater than tens of MeV or even GeV, most LPMTs will receive tens or even hundreds of PEs. For example, Fig.~\ref{fig:nPEbyPMT_500MeV_LPMTs} shows the distribution of the number of PEs (nPE) received by LPMTs for 500~MeV electrons that deposit their kinetic energies in the LS. Obviously, all LPMTs are fired in this case. If the energy deposition occurs at the edge of the detector, the nearby LPMTs will receive even more PEs. The linearity of LPMT charge reconstruction over a large charge dynamic range is a challenge and needs to be calibrated and validated in the future. Based on the experience of Daya Bay, an independent measurement system will be an effective solution~\cite{Huang-2017abb,DayaBay-2019fje}. For comparison, in the same case of 500~MeV electrons deposited their kinetic energies in the LS, there is about 45\%-60\% of SPMTs that are not triggered (Fig.~\ref{fig:fireRatio}), and most fired SPMTs only receive less than 5~PEs (Fig.~\ref{fig:nPEbyPMT_500MeV_SPMTs}) due to the fact that their photocathode areas are about 40 times smaller. Therefore, we develop an energy reconstruction algorithm using only the information from SPMTs. In addition, according to the study in~\cite{Blin-2017cwl,JUNO-2020orn}, the readout electronics of JUNO SPMTs may also exhibit non-linearity when receiving multiple hits. To minimize this effect, we will only use its firing information (fired or unfired). More details are introduced in Sec.~\ref{section:Method}
 
\section {Method of energy reconstruction}
\label{section:Method}    

\subsection{The probabilities of SPMT's firing states}
\label{subsection:FiringStates}

    For each SPMT, the number of the detected PEs obeys a Poisson distribution:
    
    \begin{equation}
    \label{eq: poisson}
    \begin{split}
        {\rm Poisson} (k_i | \mu_i)=\frac{e^{-\mu_{i}} * {\mu_{i}}^{k_{i}}}{k_i!}
    \end{split}
    \end{equation}
    
    where $k_i$ is the nPE detected by the $i^{\rm th}$ SPMT, and ${\mu_i}$ is the expected mean value of the nPE. As mentioned above, in order to reduce the dependence on charge reconstruction accuracy, the OCCUPANCY method is applied which uses only the information from SMPT's firing states (fired or unfired). Therefore, only two states of $k_i$ need to be considered: $k_i=0$ (unfired) and $k_i>0$ (fired). The probabilities of these two states can be described by Eq.~\ref{eq: trueNPEoccu0} and Eq.~\ref{eq: trueNPEoccu1}~\cite{J.W.Hardin,Zhang:2023ued}.  
    
    \begin{equation}
    \label{eq: trueNPEoccu0}
    \begin{aligned}
        P_{\rm{unfired}}(\mu_{i})={\rm Poisson} (k_i = 0 | \mu_i)=e^{-\mu_{i}}
    \end{aligned}
    \end{equation}  
    
    \begin{equation}
    \label{eq: trueNPEoccu1}
    \begin{aligned}
        P_{\rm{fired}}(\mu_{i})&={\rm Poisson} (k_i > 0 | \mu_i)\\
        &=1-P_{\rm{unfired}}(\mu_{i})=1-e^{-\mu_{i}}
    \end{aligned}
    \end{equation} 

    In real detection, $k_i$ will be smeared by fluctuation in photoelectron detection, while ${\mu_i}$ will be distorted due to the additional contribution from PMT dark count. In addition, to avoid false triggering due to electronic noise, we need to apply thresholds to each SPMT, which also affect the observation of ${\mu_i}$. In general, 0.3~PEs is chosen as the typical threshold since it can effectively handle electronic noise. Taking these effects into account, the probabilities of unfired state and fired state are: 

    \begin{equation}
    \label{eq:Punf-Pf}
    \begin{aligned}
        P_{\rm{unfired}}(\mu^{\rm{true}}_i)&=P(q_i < q_i^{\rm{threshold}} | \mu^{\rm{true}}_i)\\
        &={\rm Poisson}(k_i = 0 | \mu^{\rm{true}}_i)+P_{\rm{threLoss}}(\mu^{\rm{true}}_{i}), \\
        P_{\rm{fired}}(\mu^{\rm{true}}_i)&=P(q_i \geq q_i^{\rm{threshold}} | \mu^{\rm{true}}_i)\\
        &=1-P_{\rm{unfired}}(\mu^{\rm{true}}_i) 
    \end{aligned}
    \end{equation}

    where $q_i$ is the reconstructed charge of the $i^{\rm th}$ SPMT, $\mu^{\rm{true}}_i$ ($\mu^{\rm{true}}_i=\mu^{\rm{phy}}_i+\mu^{\rm{dn}}_i$) is the mean value of the Poisson distribution, which consists of two components: 
    
    (1) $\mu^{\rm{phy}}_i$ caused by the visible energy of physics events; 
    
    (2) $\mu^{\rm{dn}}_i$ introduced by the dark count (${\rm DR}_i$) of the $i^{\rm th}$ SPMT, and it can be calculated by $\mu^{\rm{dn}}_{i}={\rm DR}_i \times t$ in a time window of $t$; 

    $P_{\rm{threLoss}}(\mu^{\rm{true}}_{i})$ is the probability of $q_i < q_i^{\rm{threshold}}$ (0.3~PEs in this study) in the case of $k_i > 0$, calculated as following: 
    
    \begin{equation}
    \label{eq: Pfix}
    \begin{aligned}
        &P_{\rm{threLoss}}(\mu^{\rm{true}}_{i})\\
        &=\sum_{k_i=1}^{n}{[{\rm Poisson}(k_i | \mu^{\rm{true}}_{i}) \times \int_{0}^{q_i^{\rm{threshold}}}\ {\rm{Gaus}}(q_i|g_i,\sigma(g_i))\,dq_i]}
    \end{aligned}
    \end{equation}

    where $g_i=S^{\rm{gain}}_i\times{k_i}$ and $\sigma(g_i)=\sqrt{g_i} \times \sigma^{\rm{spe}}_i$, with $n$ indicates the case of multiple PEs, $S^{\rm{gain}}_i$ corresponds to the ratio between the real SPMT gain and the normal SPMT gain ($3 \times 10^6$) in JUNO, $\sigma^{\rm{spe}}_i$ denotes SPE resolution of the $i^{\rm th}$ SPMT. In real detection, $S^{\rm{gain}}_i$, $\sigma^{\rm{spe}}_i$ and ${\rm DR}_i$ can be obtained from PMT calibration. 

\subsection{Construction of the calibration map}

\begin{figure}[!htb]
        \centering          \includegraphics[width=1.\hsize]{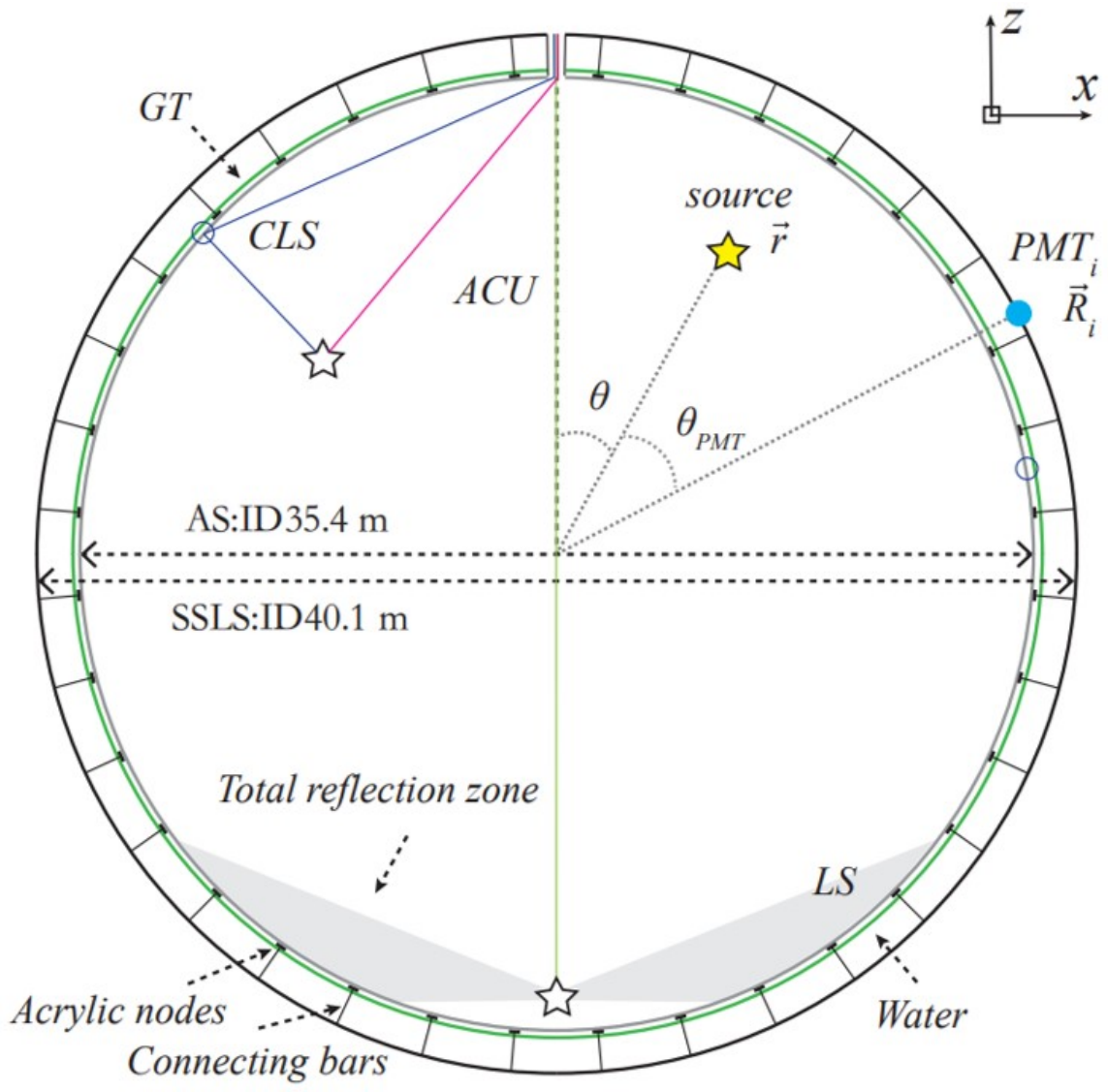}
        \caption{The individual calibration systems in the CD and their scanning regions~\cite{Huang-2021baf}.}
        \label{fig:map_strategy}
    \end{figure}

    JUNO designed a comprehensive calibration system~\cite{JUNO-PPNP} to understand the detector response, deploying multiple radioactive sources in various locations inside/outside of the CD, including Auto Calibration Unit (ACU), Cable Loop System (CLS), Guide Tube Calibration System (GTCS), and Remotely Operated Vehicle (ROV). Figure~\ref{fig:map_strategy} shows the individual calibration systems in the CD and their scanning regions. For example, ACU system scans the detector response along the central axis with multiple calibration sources, and the CLS system can scan in a 2-dimensional plane (X-Z plane) with multiple calibration sources using the central cable and side cable. The strategy of the JUNO calibration system has been developed and optimized based on Monte Carlo simulation results~\cite{JUNO-2020xtj}.

    In energy reconstruction, $\mu^{\rm{phy}}_{i}$ directly corresponds to the visible energy of an event in the detector, which is the basis of the energy reconstruction in our method. Assuming that a calibration source is loaded at the location $ \overrightarrow{r}(r,\theta,\phi=0)$ in the central detector, the mean value of visible-energy-induced PEs for the $i^{\rm th}$ SPMT is $\mu^{\rm{phy\_source}}_{i}$, which corresponds to the visible energy (denoted as $E^{\rm{source}}$) of the calibration source. Then for an event depositing its energy at the same location, the relationship between the event's visible energy $E_{\rm{vis}}$ and $\mu^{\rm{phy}}_{i}$ for the $i^{\rm th}$ SPMT can be described as follows:
   
    \begin{equation}
    \label{eq:mu}
    \begin{split}
        \mu^{\rm{phy}}_{i}=\frac{E_{\rm{vis}}}{E^{\rm{source}}} \times \mu^{\rm{phy\_source}}_{i} 
    \end{split}
    \end{equation}

    It should be noted that $\mu^{\rm{phy}}_{i}$ is not only related to the visible energy and the position of the event, but also to the relative position ($\theta_{\rm{SPMT}}$) of the event and the $i^{\rm th}$ SPMT. This relationship can be determined using the calibration data, which means constructing a calibration map. In Sec.~\ref{subsection:likelihood}, the maximum likelihood method will be adopted to reconstruct the visible energy by estimating $\mu^{\rm{phy}}_{i}$ via the calibration map and invoking the firing states of 25600 SPMTs from data. Next, the construction of the calibration map will be introduced.  
      
    According to the calibration strategy in JUNO~\cite{JUNO-2020xtj}, this study uses the $^{68}$Ge source (positron source, with $E^{\rm{source}}=1.022$~MeV) to calibrate the X-Z plane with assistance from both ACU and CLS systems across 227 positions. Using the JUNO offline software~\cite{Li_2017zku,Li-2018fny,Lin:2022htc}, 10000 $^{68}$Ge events are generated respectively for each calibration location on the X-Z calibration plane. Realistic detector geometry is employed for all of these samples, and the optical parameters of LS are implemented based on precise measurements~\cite{Zhou-2015gwa, Gao-2013pua, Wurm-2010ad, Zhang-2020mqz, Ding-2015sys, Buck-2015jxa, OKeeffe-2011dex}. Comprehensive optical processes are simulated using Geant4~\cite{GEANT4-2002zbu}. Furthermore, the official electronic simulation (including SPMT's charge smearing, transit time spread and dark noise, etc, which are referenced in the measurement~\cite{Cao-2021wrq}) and charge reconstruction are also applied in this study.

    In the calibration, when the $^{68}$Ge source is loaded at one of the 227 planned locations, the probability of the unfired state for the $i^{\rm th}$ SPMT can be estimated as Eq.~\ref{eq:P0-singlePMT}. Where $N_{q_{i} < q_{i}^{\rm{threshold}}}$ corresponds to the number of events with $q_{i} < q_{i}^{\rm{threshold}}$, and $N_{\rm{total}}$ is the total number of events for the calibration sample, and in this study $N_{\rm{total}}=10000$. According to Eq.~\ref{eq:Punf-Pf}, for convenience, we use an effective mean value of detected PEs ($\mu^{\rm{det\_source}}_i$) for the $i^{\rm th}$ SPMT in Eq.~\ref{eq:P0mu-singlePMT}. Then $\mu^{\rm{det\_source}}_i$ can be estimated by Eq.~\ref{eq:mu-singlePMT} using the calibration data. Obviously, this calculation requires that the $i^{\rm th}$ SPMT is fired not for all 10000 events in the calibration sample ($^{68}$Ge), otherwise this method is no longer applicable. Due to the small visible energy of $^{68}$Ge and the fact that even the most marginal of the 227 calibration positions is about 2 meters away from its neighboring SPMTs, the above extreme scenario is extremely unlikely.
    
    \begin{equation}
    \label{eq:P0-singlePMT}
    \begin{aligned}
        P_{\rm{unfired}}(\mu^{\rm{true\_source}}_i)&=P(q_{i} < q_{i}^{\rm{threshold}} | \mu^{\rm{true\_source}}_i)\\
        &=\frac{N_{q_{i} < q_{i}^{\rm{threshold}}}}{N_{\rm{total}}}
    \end{aligned}
    \end{equation}

    \begin{equation}
    \label{eq:P0mu-singlePMT}
    \begin{aligned}
        P_{\rm{unfired}}(\mu^{\rm{true\_source}}_i)&={\rm Poisson}(k_i = 0 | \mu^{\rm{det\_source}}_i)\\
        &=e^{-\mu^{\rm{det\_source}}_i} 
    \end{aligned}
    \end{equation}

    \begin{equation}
    \label{eq:mu-singlePMT}
    \begin{split}
        \mu^{\rm{det\_source}}_{i}&=-\ln{P_{\rm{unfired}}}\\
        &=-\ln{\frac{N_{q_i < q_{i}^{\rm{threshold}}}}{N_{\rm{total}}}}
    \end{split}
    \end{equation}

\begin{figure}[!htb]
        \centering  \includegraphics[width=0.9\hsize]{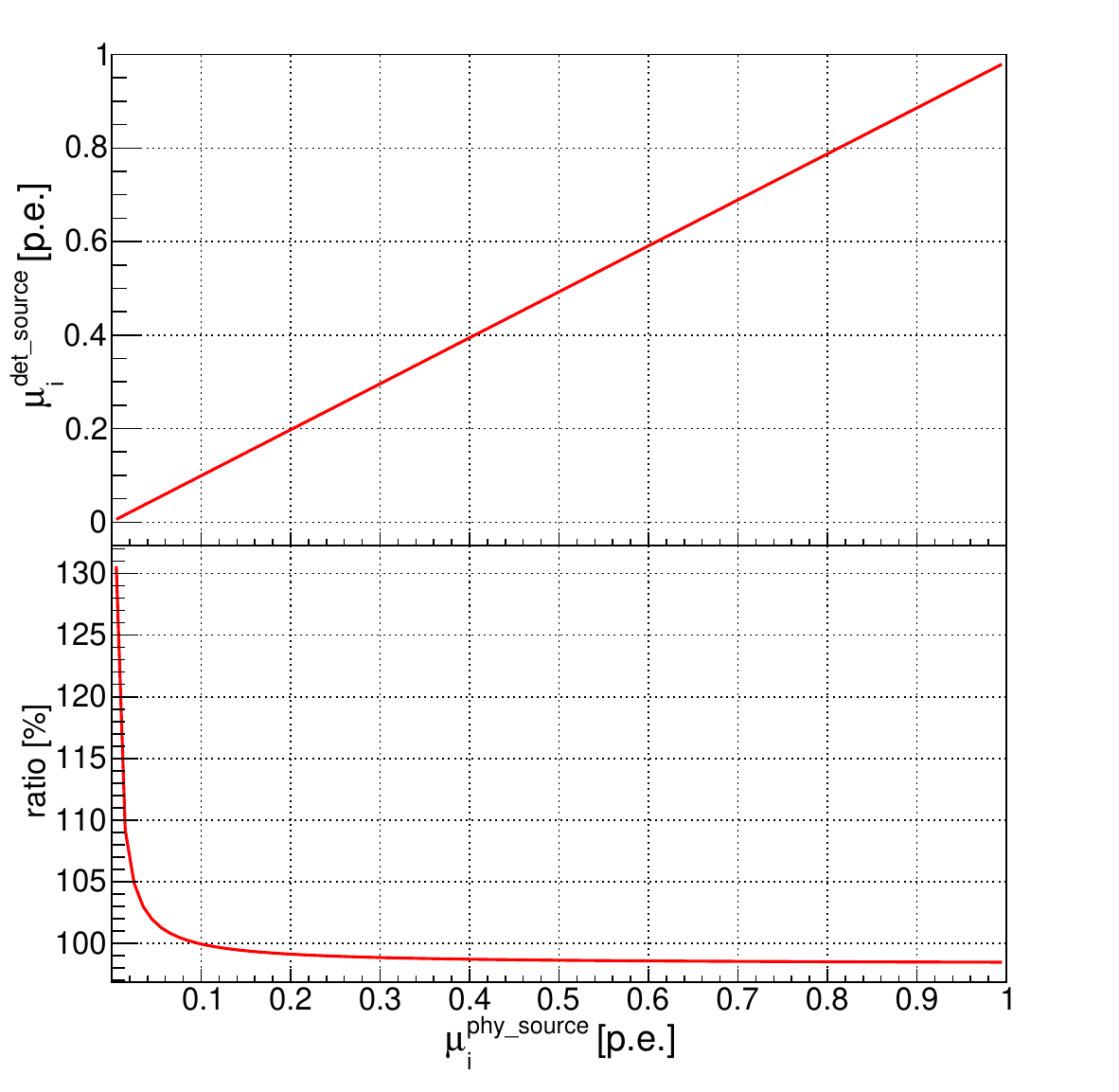}
        \caption{The relationship between $\mu^{\rm{det\_source}}_i$ and $\mu^{\rm{phy\_source}}_i$ for a SPMT whose spe resolution and dark count rate are 30\% and 1~kHz, respectively; and the readout window is set to 1000~ns. The bottom panel shows the ratio of $\mu^{\rm{det\_source}}_i$/$\mu^{\rm{det\_source}}_i$.}
        \label{fig:mufixed}
    \end{figure}

    It should be noted that $\mu^{\rm{phy\_source}}_i$ is the value required for energy reconstruction (Eq.~\ref{eq:mu}), while $\mu^{\rm{det\_source}}_i$ includes the contributions from visible energy, PMT dark count, charge smearing and the threshold effect. According to Eq.~\ref{eq:Punf-Pf}, Eq.~\ref{eq:P0mu-singlePMT} and PMT parameters ($S^{\rm{gain}}_i$, $\sigma^{\rm{spe}}_i$ and ${\rm DR}_i$) from PMT calibration, we can find the relationship between $\mu^{\rm{phy\_source}}_i$ and $\mu^{\rm{det\_source}}_i$. Figure~\ref{fig:mufixed} shows an example of an SPMT, whose spe resolution and dark count rate are 30\% and 1~kHz, respectively. The readout window is 1000~ns. It appears that dark counts dominate $\mu^{\rm{det\_source}}_i$ for small $\mu^{\rm{phy\_source}}_i$, while the combined effect of dark count, smearing and threshold remains stable at around 2\% for the given setting as $\mu^{\rm{phy\_source}}_i$ increases. 

\begin{figure}[!htbp]
    \centering
    \subfigure[$\theta_{SPMT}$ in [$18^{\circ}$,$18.125^{\circ}$)]
    {
        \label{fig:map144nointer}
        \includegraphics[width=0.8\hsize]{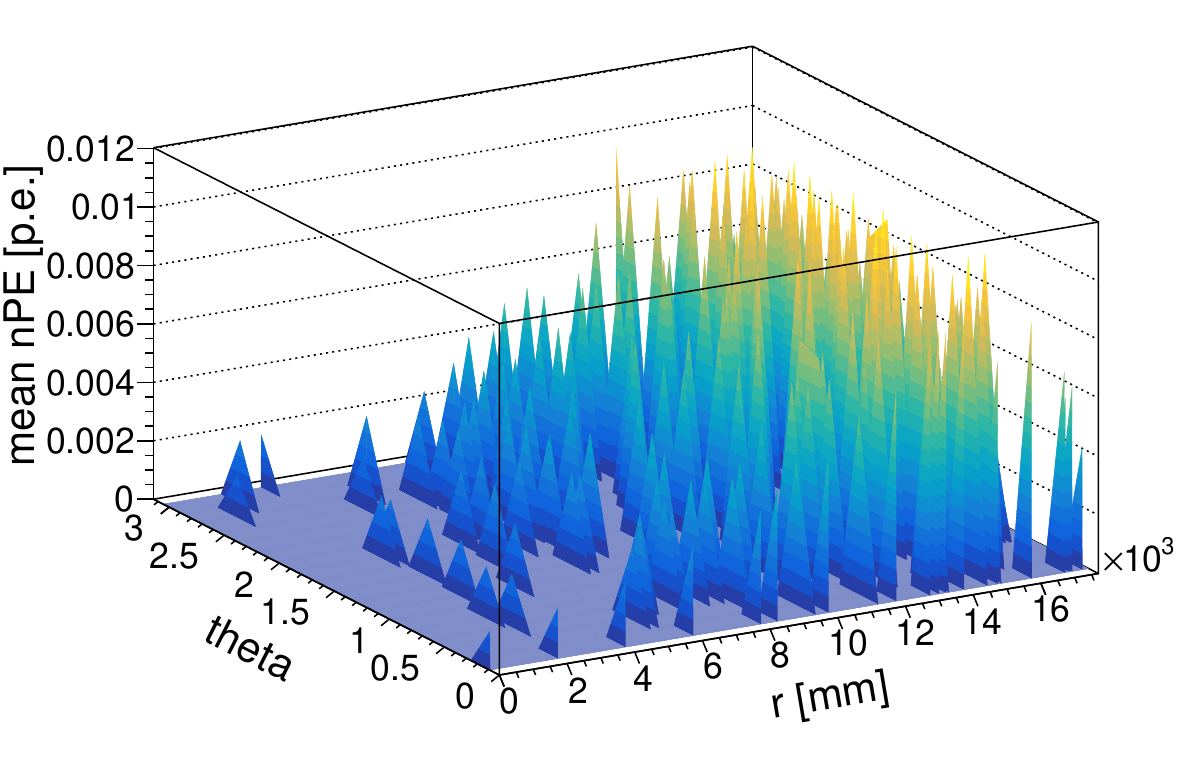}
    }
    \subfigure[$\theta_{SPMT}$ in [$18^{\circ}$,$18.125^{\circ}$)]
    {
        \label{fig:map144inter}
        \includegraphics[width=0.8\hsize]{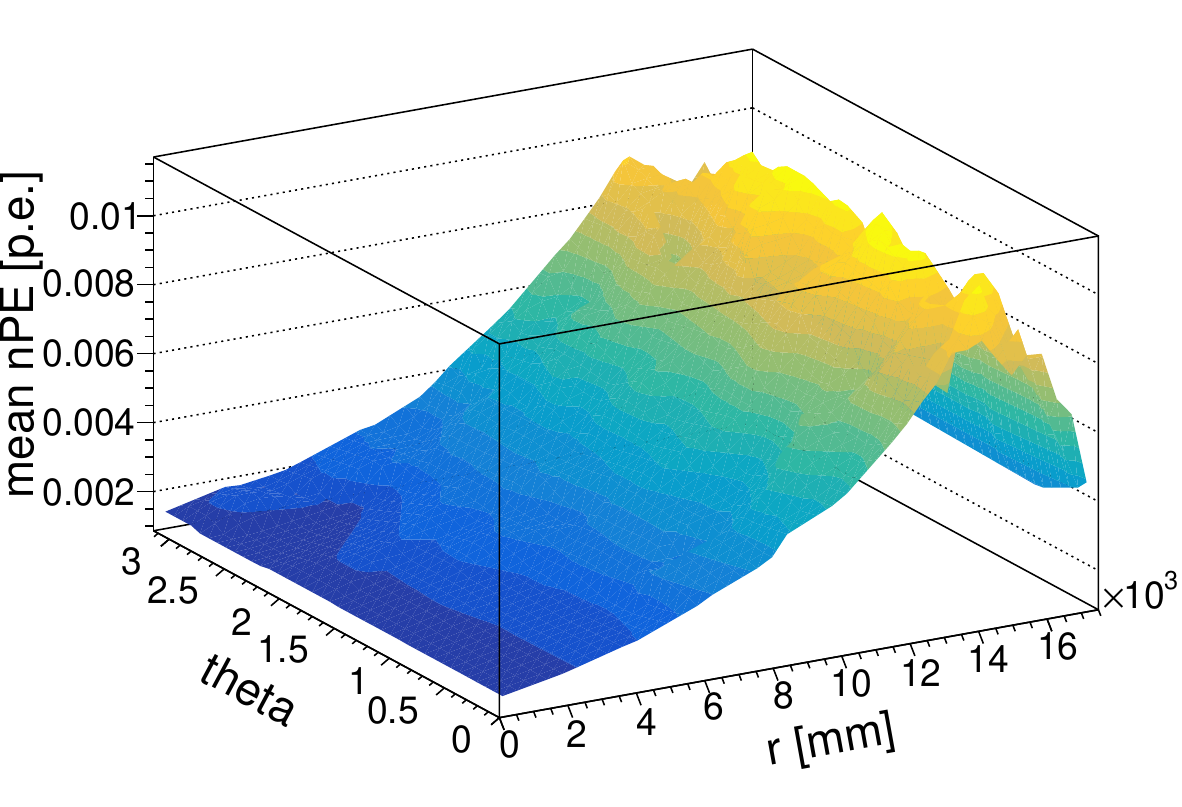}
    }
    \caption{Example of calibration map for one $\theta_{SPMT}$ group whose $\theta_{PMT}$ values from $18^{\circ}$ to $18.125^{\circ}$.}
    \label{3dmap_example}
    \end{figure}

    Considering that the CD has good symmetry and the SPMTs with the same relative position with respect to the calibration source exhibit similar responses, in order to enhance the accuracy, we further group and combine SPMTs with similar $\theta_{\rm{SPMT}}$ values in the calculation of $\mu^{\rm{phy\_source}}_i$. In this work, $\theta_{\rm{SPMT}}$ is divided into 1440 groups from $0^{\circ}$ to $180^{\circ}$, with $0.125^{\circ}$ per group. The same approach has been successfully verified and applied in~\cite{Huang-2022zum}. As a result, for each $^{68}$Ge source location $\overrightarrow{r}(r,\theta,\phi=0)$, SPMTs in the same $\theta_{\rm{SPMT}}$ group have similar values of $\mu^{\rm{phy\_source}}_i$, the average of which is denoted as $\mu^{\rm{phy\_source}}(\overrightarrow{r},\theta_{\rm{SPMT}})$. The calibration map can be constructed after $^{68}$Ge scans 227 locations on the X-Z plane and all $\mu^{\rm{phy\_source}}(\overrightarrow{r},\theta_{\rm{SPMT}})$ calculated. Considering the calibration performance and time consumption in JUNO, there are only about 227 calibration points available in the current calibration strategy. So it's necessary to apply interpolation for the remaining positions. Figure~\ref{fig:map144nointer} and Fig.~\ref{fig:map144inter} show examples of the calibration map before and after interpolation, respectively.

    The following is a brief summary of the main steps in constructing the calibration map.

    (1) For $^{68}$Ge loading at a location $\overrightarrow{r}(r,\theta,\phi=0)$, calculate the effective mean value of the detected PEs ($\mu^{\rm{det\_source}}_i$) for each SPMTs using Eq.~\ref{eq:mu-singlePMT};

    (2) Calculate $\mu^{\rm{phy\_source}}_i$ by correcting PMT dark counts and threshold effect, Fig.~\ref{fig:mufixed} shows an example of the relationship between $\mu^{\rm{det\_source}}_i$ and $\mu^{\rm{phy\_source}}_i$ for a SPMT; 

    (3) Calculate $\mu^{\rm{phy\_source}}(\overrightarrow{r},\theta_{\rm{SPMT}})$, which is the average value of $\mu^{\rm{phy\_source}}_i$ for each $\theta_{\rm{SPMT}}$ group;

    (4) Repeat the above steps for calibration data at all locations;

    (5) Apply interpolation to the remaining positions;

    (6) For a given $^{68}$Ge location $\overrightarrow{r}(r,\theta,\phi=0)$, the $\mu^{\rm{phy\_source}}(\overrightarrow{r},\theta_{\rm{SPMT}})$ value corresponding to each SPMTs can be obtained;

    The calibration map is generated using the calibration data on the X-Z plane ($\phi=0$) by considering that the detector exhibits good symmetry along the $\phi$ direction~\cite{JUNO-2020xtj}. The $\phi$ symmetry is reliable for events located within 16~m, according to a detailed study in~\cite{Huang-2021baf}. However, the $\phi$ dependence can no longer be ignored for the region near the edge due to the shadowing effect of numerous acrylic nodes. In the future, the $\phi$ symmetry needs to be checked and validated using real data, and it can be corrected if necessary.

\subsection{Construction of maximum likelihood function}
\label{subsection:likelihood}

    To reconstruct the visible energy of a cluster-like event whose energy-deposit center is known in the detector, a likelihood function can be constructed as follows:

    \begin{equation}
    \label{eq:likelihood0}
        \mathcal{L}=\prod \limits_{1}^{N_{\rm{unfired}}}P_{\rm{unfired}}(\mu^{\rm{phy}}_i)\prod \limits_{1}^{N_{\rm{fired}}}P_{\rm{fired}}(\mu^{\rm{phy}}_i)
    \end{equation}
    
    where $N=N_{\rm{unfired}}+N_{\rm{fired}}=25600$, $N_{\rm{unfired}}$ and $N_{\rm{fired}}$ correspond to the number of SPMTs with $q_i<0.3$~PEs and $q_i\geq0.3$~PEs, respectively. $P_{\rm{unfired}}(\mu^{\rm{phy}}_i)$ and $P_{\rm{fired}}(\mu^{\rm{phy}}_i)$ are the probabilities of unfired state and fired state, which can be calculated using Eq.~\ref{eq:Punf-Pf}. In the calculation, $\mu^{\rm{phy}}_i$ of each SPMT can be estimated by considering the relationship in Eq.~\ref{eq:mu} and invoking the $\mu^{\rm{phy\_source}}(\overrightarrow{r},\theta_{\rm{SPMT}})$ value from the calibration map. In addition, differences in quantum efficiency (QE) between individual SPMTs in the same $\theta_{\rm{SPMT}}$ group should be considered. As a result, $\mu^{\rm{phy\_source}}_i$ needs a correction and it can be calculated in the following:

    \begin{equation}
    \label{eq:calculateMu}
    \begin{split}
        \mu^{\rm{phy\_source}}_i=\frac{QE_i}{\frac{1}{m}\sum_{j=0}^{m}{QE_j}} \times \mu^{\rm{phy\_source}}(\overrightarrow{r},\theta_{\rm{SPMT}})
    \end{split}
    \end{equation}

    where $QE_i$ is the QE of the $i^{\rm th}$ SPMT, $m$ is the number of SPMTs in the $\theta_{\rm{SPMT}}$ group which is classified by the $\theta_{\rm{SPMT}}$ value, and $QE_j$ is the QE of the $j^{\rm th}$ SPMT in this group. 

\begin{figure}[!htb]
        \centering \includegraphics[width=0.9\hsize]{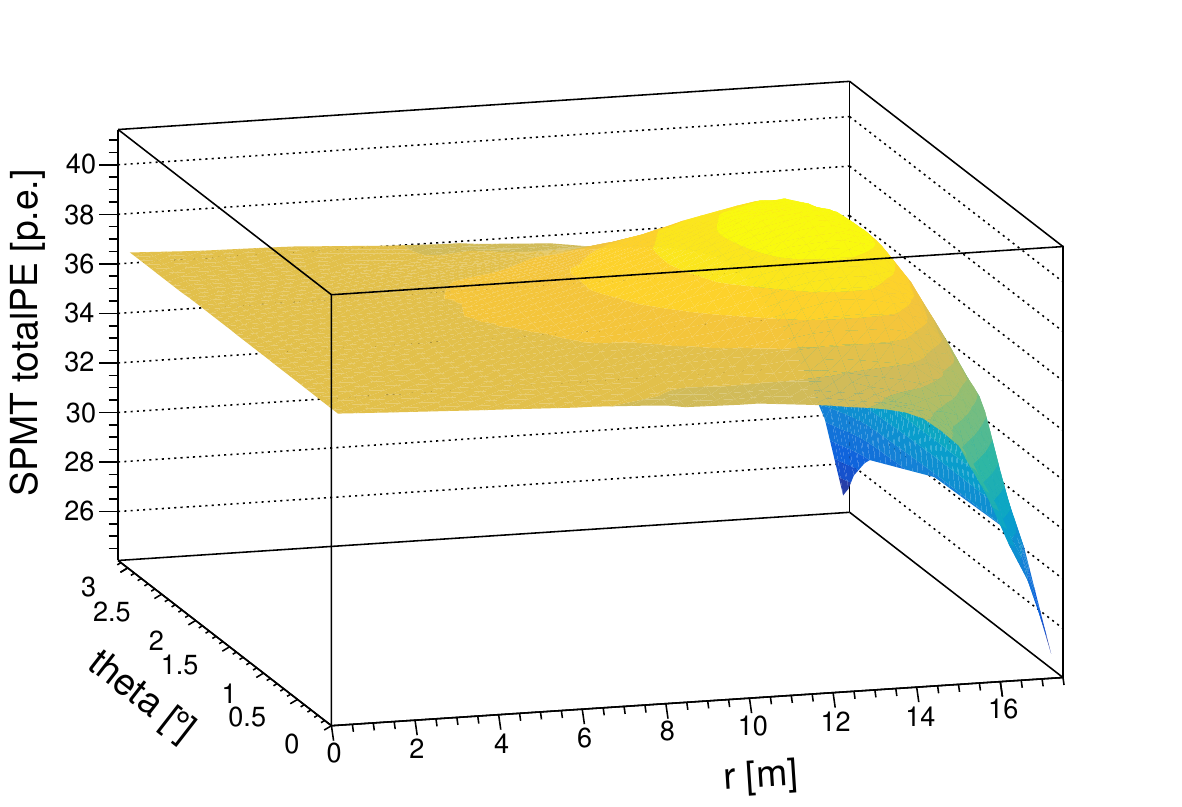}
        \caption{The total number of PEs observed by 25600 SPMTs for $^{68}$Ge source located at different positions.}
        \label{fig:scaleMap}
    \end{figure}

    Next, the ROOT's minimization class TMinuit2Minimizer~\cite{Brun-1997pa,James-1975dr,Hatlo-2005cj} is used for minimizing $-\ln{\mathcal{L}}$. In minimization, the visible energy $E_{\rm{vis}}$ of the cluster-like event (whose energy-deposit center is known and will be introduced in Sec.~\ref{subsection:VertexReconstruction}) is the parameter to be determined, and its initial value ($E_{\rm{initial}}$) can be estimated as Eq.~\ref{eq:es} using the total number of PEs ($totalPE$) of all SPMTs to reduce the reconstruction time. $totalPE^{\rm{source}}(r)$ is $totalPE$ observed by 25600 SPMTs for the $^{68}$Ge source located at different positions (Fig.~\ref{fig:scaleMap}). Comparing the value of $totalPE^{\rm{source}}(r)$ at the center of CD and at around $r=15$~m, there is about 7\% non-uniformity introduced by the reception of PMTs and the optical attenuation effect in the process of photon transmission. As for the decreasing radius larger than $\sim$15.6 m, it is mainly caused by the total reflection effect and the shadowing effect.

    \begin{equation}
    \label{eq:es}
    \begin{split}
        E_{\rm{initial}} = \frac{totalPE(r)}{totalPE^{\rm{source}}(r)} \times E^{\rm{source}}
    \end{split}
    \end{equation}
    
    After minimization, TMinuit returns the predicted visible energy value as the result of energy reconstruction. Furthermore, the response difference caused by the spatial scale of energy deposition is investigated in Sec.~\ref{subsection:comparisonCluster} and it is found that has little influence on our analysis. 
    

\subsection{Comparison of cluster-like event and point-like event}
\label{subsection:comparisonCluster}

    $^{68}$Ge is a positron source. The positron annihilated in the source capsule (stainless steel + PTFE) with a pair of 0.511~MeV $\gamma$s emitted. Most of the energy of $\gamma$s is deposited within $\sim$30~cm in the LS, so $^{68}$Ge is not strictly a point-like source, but a cluster-like source similar to the high-energy electrons described earlier in Sec.~\ref{section:Introduction}. In this paper, to determine the location-dependent $\mu_i^{\rm{phy\_source}}$ from calibration data and carry out energy reconstruction, we construct a calibration map using the approximate point-like $^{68}$Ge source (cluster size in $\sim$30~cm). It is required to investigate if this calibration map can be applied to higher energy electrons with larger cluster size (up to several meters), by assuming they are point-like sources with their energy-deposit centers as the source positions.

    According to Eq.~\ref{eq:mu}, for an event with a known position in the LS, with $E^{\rm{source}}$ and $\mu_i^{\rm{phy\_source}}$ obtained from the calibration map, as shown in Eq.~\ref{eq:mu_sum}, the sum of all SPMT's $\mu_i^{\rm{phy}}$ is directly proportional to the visible energy of the event. Therefore, any bias present in the sum of all SPMT's $\mu_i^{\rm{phy}}$ indicates the deviation of the visible energy.

    \begin{equation}
    \label{eq:mu_sum}
    \begin{split}
        \sum_{i = 1}^{N}{\mu^{\rm{phy}}_{i}}=\frac{E}{E^{\rm{source}}} \times \sum_{i = 1}^{N}{\mu_i^{\rm{phy\_source}}} 
    \end{split}
    \end{equation}

    In energy reconstruction, our algorithm assumes that all energy deposition is equivalent to occurring at the energy-deposit center. However, in real detection, $\mu^{\rm{phy}}_{i}$ of the $i^{\rm th}$ SPMTs is contributed by the cumulative effect of each secondary energy deposition. To estimate the potential bias caused by the spatial scale of energy deposition for high-energy events, the sum of all SPMT's $\mu_i^{\rm{phy}}$ (Eq.~\ref{eq:mu_sum}) is adopted using different strategies. As shown in Eq.~\ref{eq:mu_method1} and Eq.~\ref{eq:mu_method2}, they are calculated as point-like events and cluster-like events, respectively.

    \begin{equation}
    \label{eq:mu_method1}
    \begin{aligned}
        &\sum_{i = 1}^{N}{\mu^{\rm{phy}}_{i}}|_{\rm{point-like}}=\\
        &\frac{1}{E^{\rm{source}}} \times \sum_{i = 1}^{N}{ \sum_{\alpha = 1}^{N_{E}}{E_{\alpha} \times \mu^{\rm{phy\_source}}_{i}(\overrightarrow{r})}}
    \end{aligned}
    \end{equation}

    \begin{equation}
    \label{eq:mu_method2}
    \begin{aligned}
        &\sum_{i = 1}^{N}{\mu^{\rm{phy}}_{i}}|_{\rm{cluster-like}}=\\
        &\frac{1}{E^{\rm{source}}} \times \sum_{i = 1}^{N}{ \sum_{\alpha = 1}^{N_{E}}{E_{\alpha} \times \mu^{\rm{phy\_source}}_{i}(\overrightarrow{r_\alpha})}}
    \end{aligned}
    \end{equation}

    \begin{equation}
    \label{eq:mu_ratio}
    \begin{aligned}
        Ratio=\frac{\sum_{i = 1}^{N}{\mu^{\rm{phy}}_{i}}|_{\rm{cluster-like}}}{\sum_{i = 1}^{N}{\mu^{\rm{phy}}_{i}}|_{\rm{point-like}}}
    \end{aligned}
    \end{equation}

\begin{figure}[!htb]
        \centering \includegraphics[width=0.9\hsize]{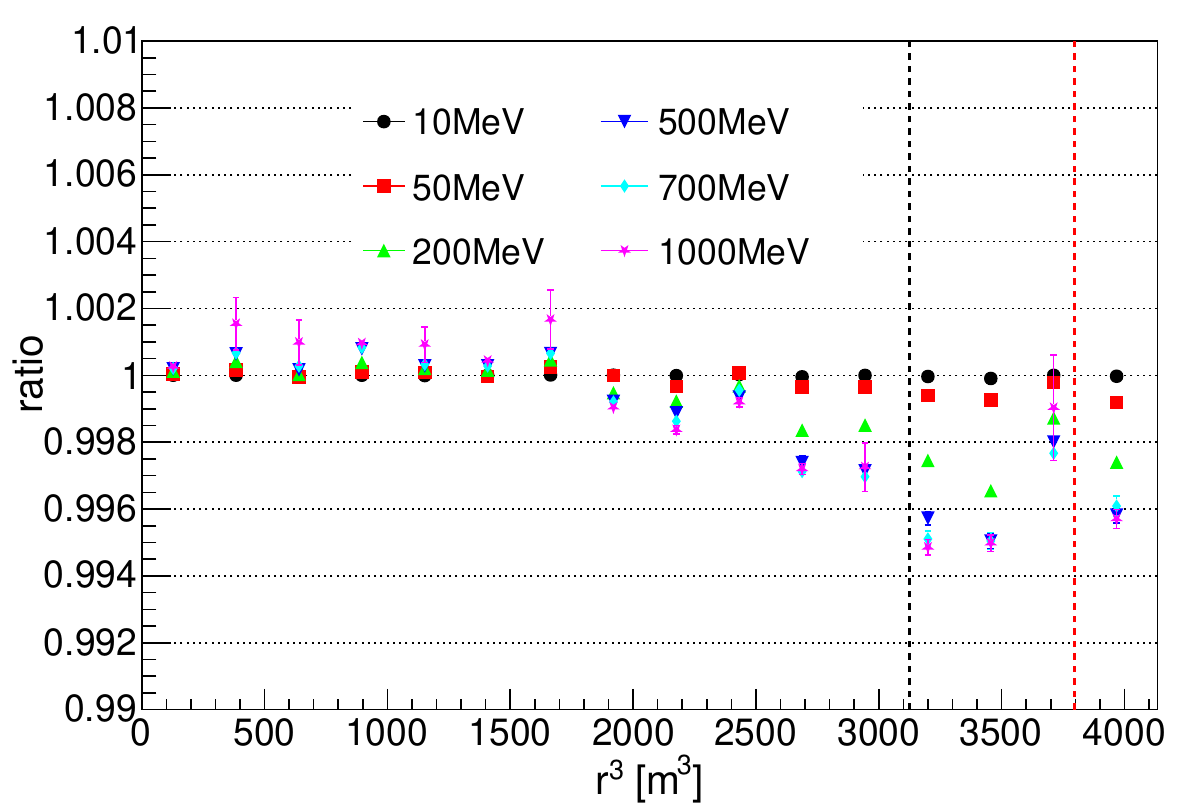}
        \caption{Comparison between the calculation results of point-like and cluster-like treatment. The ratio, defined in Eq.~\ref{eq:mu_ratio}, was found to be close to 1 for electron samples with different kinetic energies at various positions.}
        \label{fig:stepratio}
    \end{figure}

    The official simulation software of JUNO is applied to generate electron samples with different kinetic energies, and the details of energy deposition ($E_{\alpha}$ and $\overrightarrow{r_{\alpha}}$) are recorded for the calculation of Eq.~\ref{eq:mu_method1} and Eq.~\ref{eq:mu_method2}. Then we compare the calculation results as shown in Eq.~\ref{eq:mu_ratio} and Fig.~\ref{fig:stepratio}. It was found that the ratio is close to 1 with the increase of $r^3$, and it has a small bias ($\leq 0.5\%$) at the edge for electrons whose kinetic energies are larger than 200~MeV. This result indicates that our algorithm and reconstruction strategy is applicable to the energy reconstruction of high-energy events that have large spatial scales of energy deposition. 

\section{Reconstruction result} 
\label{section:Result}

    In this section, the JUNO offline software is used to simulate electrons uniformly distributed in the CD with different kinetic energies (10, 20, 50, 100, 200, 350, 500, 700, 1000, and 2000~MeV) as the MC sample to validate the reconstruction algorithm. The MC samples are generated after a full-chain simulation in JUNO by including realistic detector geometry, comprehensive physical interaction processes, and optical transmission processes, official electronic simulation, and charge reconstruction. A 16~m radius cut is applied to avoid energy leakage for high-energy events and the total reflection effect at the edge of the detector. Finally, there are about 10000 events for each electron sample.

\subsection {Reconstruction of energy-deposit center} 
\label{subsection:VertexReconstruction}
    As introduced in Sec.~\ref{subsection:likelihood}, the event's energy-deposit center is required to reconstruct the visible energy. So, before energy reconstruction, it's necessary to reconstruct the energy-deposit center. Compared to the vertex reconstruction of point-like events, the energy-deposit center reconstruction of high-energy cluster-like events faces greater dispersion (Sec.~\ref{section:Introduction}). After investigation, it is found that the time-based algorithm developed and verified by~\cite{Li-2021oos} is suitable and can be applied in our analysis.

    \begin{figure*}[!htb]
    \centering 
    \subfigure[Reconstruction bias of $r$]
    {
    \label{Fig-sub1-recr}
    \includegraphics[width=0.32\hsize]{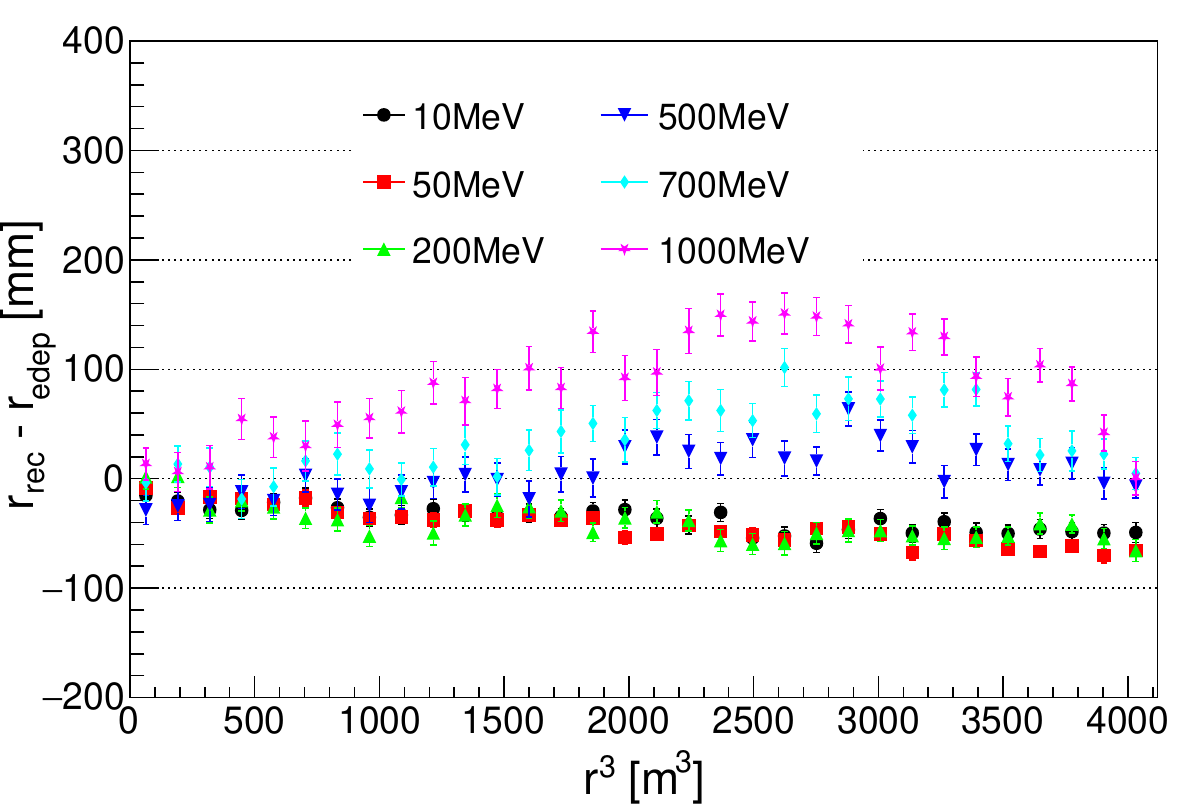}
    }
    \subfigure[Reconstruction bias of $\theta$ direction]
    {
    \label{Fig-sub2-rect}
    \includegraphics[width=0.32\hsize]{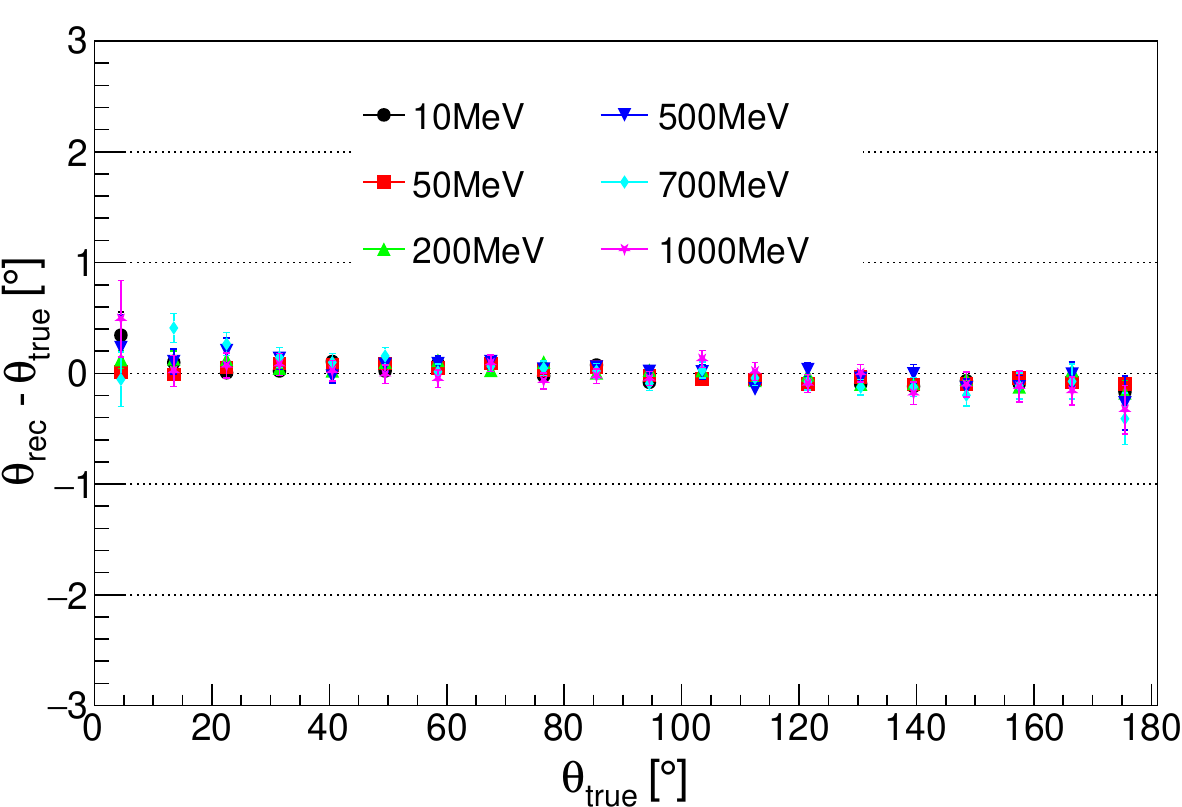}
    }
    \subfigure[Reconstruction bias of $\phi$ direction]
    {
    \label{Fig-sub3-recp}
    \includegraphics[width=0.32\hsize]{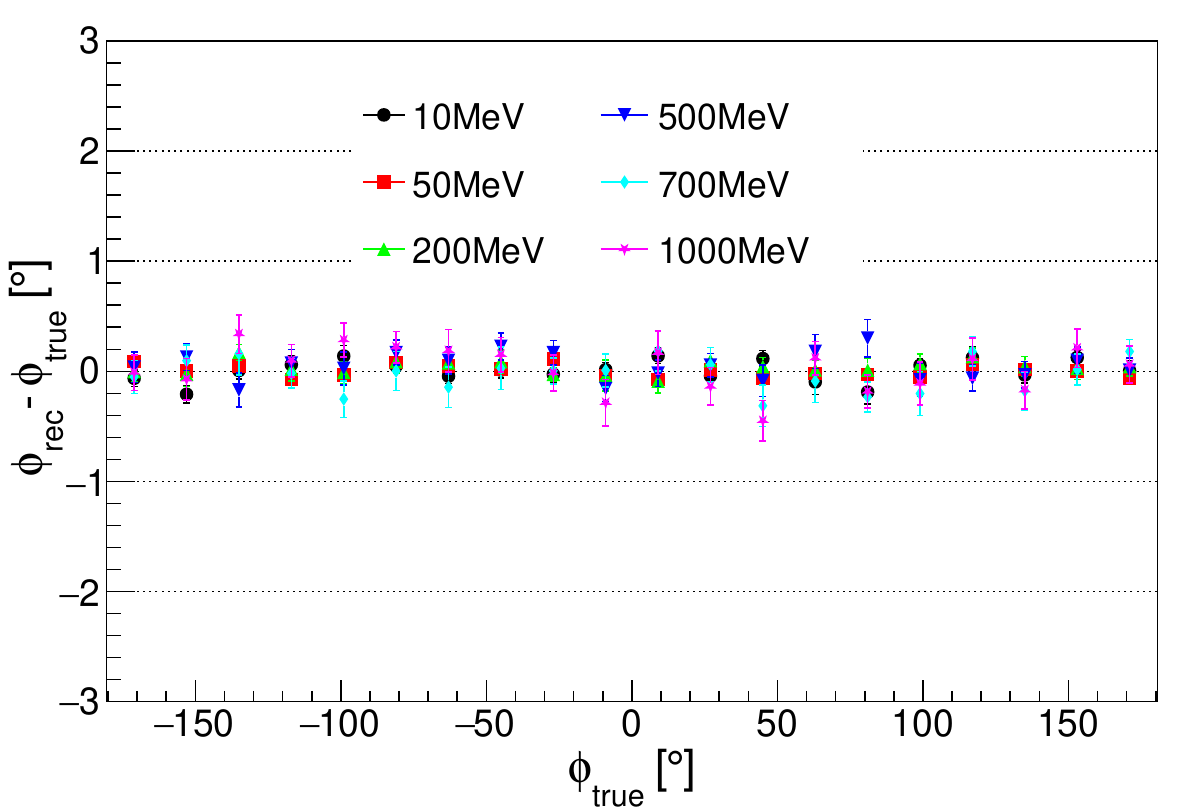}
    }
    \caption{Reconstruction bias of electron's energy-deposit center.}
    \label{Fig:vertexrec}
    \end{figure*}

    The time-based algorithm uses the distribution of time-of-flight (${\rm t.o.f.}$) corrected time $\Delta t$ (Eq.~\ref{eq:ptf}) of an event to reconstruct its vertex and $t_0$ (event time). The principle of the time-based algorithm is that the $\Delta t$ distribution is independent of the event vertex after applying the time-of-flight correction. In this paper, we apply it to the reconstruction of the energy-deposit center. 
    
    \begin{equation}
    \label{eq:ptf}
    \begin{split}
        \Delta t_i=t_i-{\rm t.o.f.}_i
    \end{split}
    \end{equation}
    
\begin{figure*}[t]
    \centering 
    \subfigure[Reconstruction resolution of $r$]
    {
    \label{Fig.sub.1 recr res}
    \includegraphics[width=0.32\hsize]{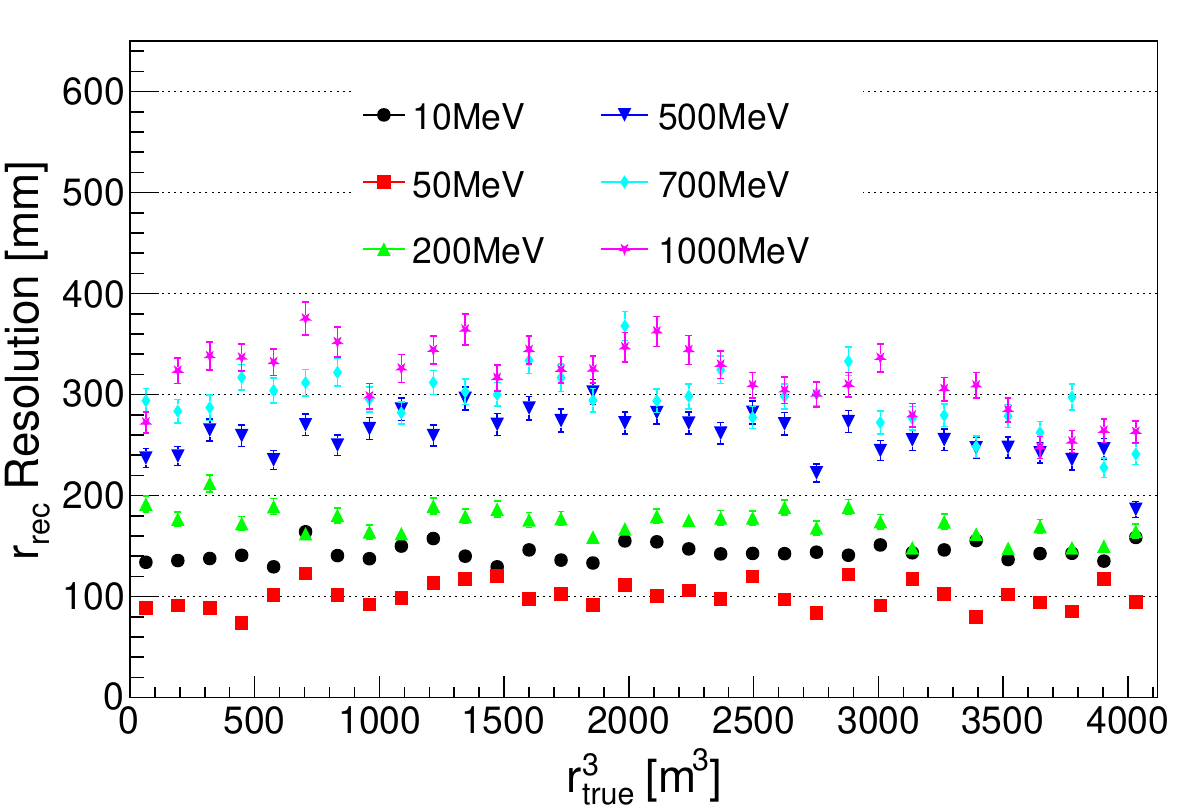}
    }
    \subfigure[Reconstruction resolution of $\theta$]
    {
    \label{Fig.sub.2 rect res}
    \includegraphics[width=0.32\hsize]{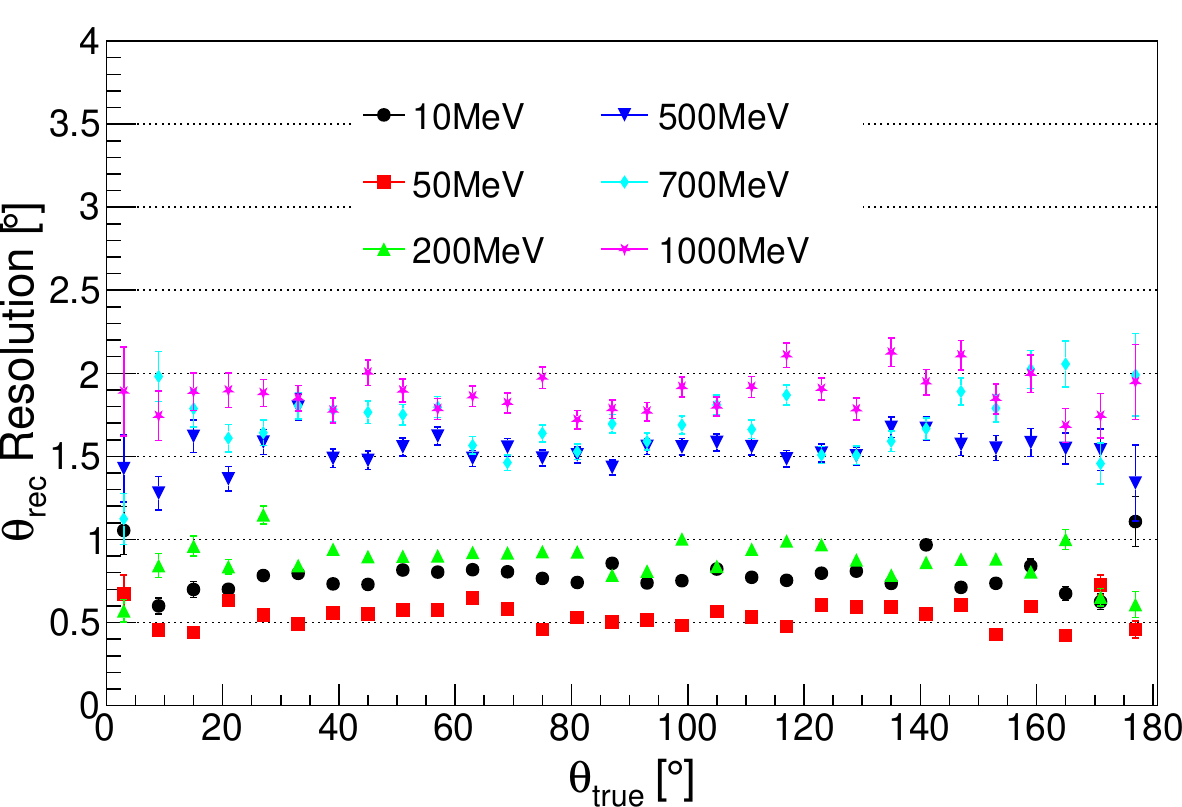}
    }
    \subfigure[Reconstruction resolution of $\phi$]
    {
    \label{Fig.sub.3 recp res}
    \includegraphics[width=0.32\hsize]{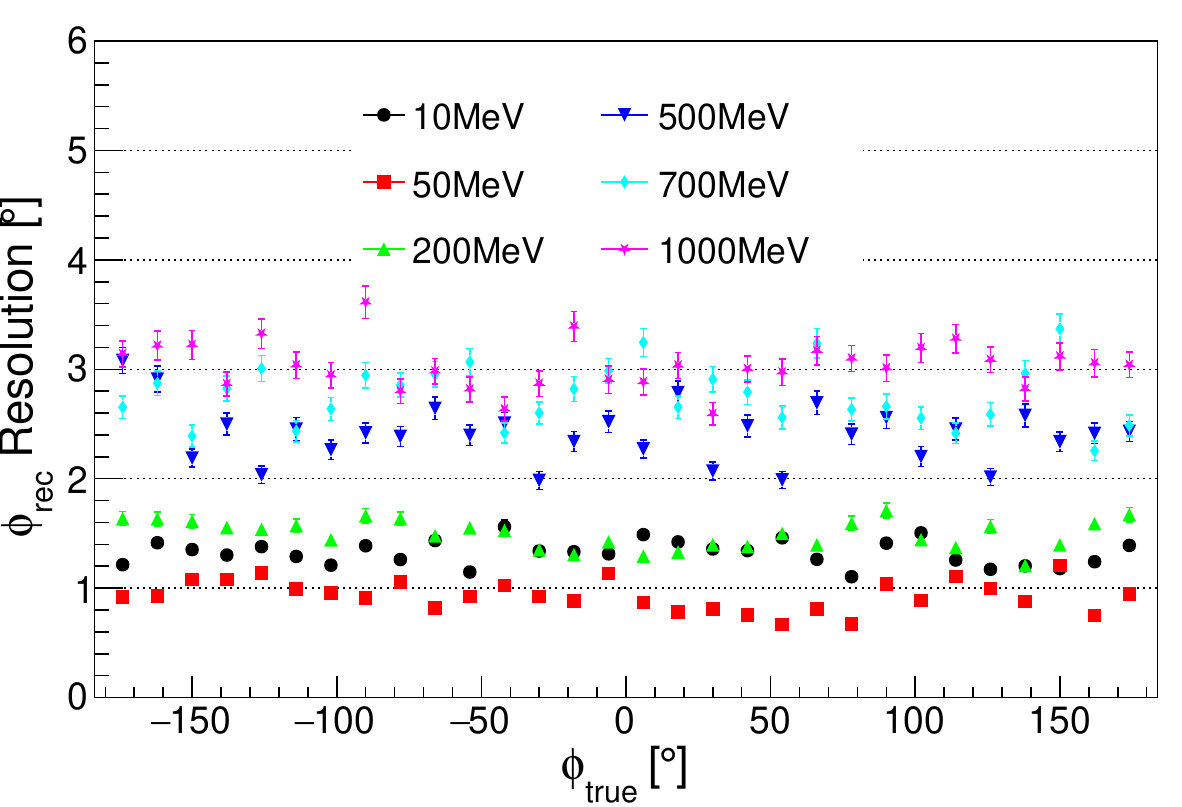}
    }
    \caption{Reconstruction resolution of electron's energy-deposit center.}
    \label{Fig:vertexrec-res}
    \end{figure*}

    In Eq.~\ref{eq:ptf}, $t_i$ is the first hit time of the $i^{\rm th}$ SPMT and ${\rm t.o.f.}_i$ is the time of flight from the energy-deposit center to the $i^{\rm th}$ SPMT. In the calculation of ${\rm t.o.f.}_i$, the optical path length includes both the length in the LS and in the water. Then a correction vector will be constructed and minimized by iterating the energy-deposit center. More details can be found in \cite{Li-2021oos}. Finally, the reconstructed energy-deposit center can be obtained and Fig.\ref{Fig:vertexrec} and Fig.~\ref{Fig:vertexrec-res} show the performance. The reconstruction biases of $\theta$ (Fig.~\ref{Fig-sub2-rect}) and $\phi$ (Fig.~\ref{Fig-sub3-recp}) remain small and stable as the energy increases, while the reconstruction bias of $r$ (Fig.~\ref{Fig-sub1-recr}) gradually increases with energy. However, it can still be controlled within 150~mm at 1~GeV. Considering that the cluster size could be several meters for a 1~GeV electron, this bias is still acceptable. And the effect of the deviation on energy reconstruction will be discussed later. On the other hand, the reconstruction bias of $R$ tends to decrease at the edge of the detector compared to other regions, mainly due to energy leakage near the edge, especially for high-energy events.

    In Fig.~\ref{Fig:vertexrec-res}, it can be found that the reconstruction resolutions of $r$, $\theta$ and $\phi$ increase with electron energy between 50~MeV and 1~GeV. For example, the resolutions of $r$, $\theta$ and $\phi$ are about 100~mm, $0.5^\circ$ and $1.0^\circ$ for 50~MeV electrons, respectively, while the resolutions of $R$, $\theta$ and $\phi$ are about 340~mm, $1.8^\circ$ and $3.0^\circ$ for 1~GeV electrons, respectively. This effect is mainly due to the greater dispersion of the energy deposition of high-energy electrons. The resolutions of 10~MeV electrons are slightly larger than the resolutions of 50~MeV electron. This is because the hit number of SPMTs is small ($\sim$400~PEs for 10~MeV) and less information is available for the reconstruction using the time-based algorithm.

\subsection {Energy reconstruction performance} 
\label{subsection:Performance}

\begin{figure*}[!htb]
    \centering  
    \subfigure[10~MeV]
    {
    \label{Fig.sub.1 calib recv}
    \includegraphics[width=0.32\hsize]{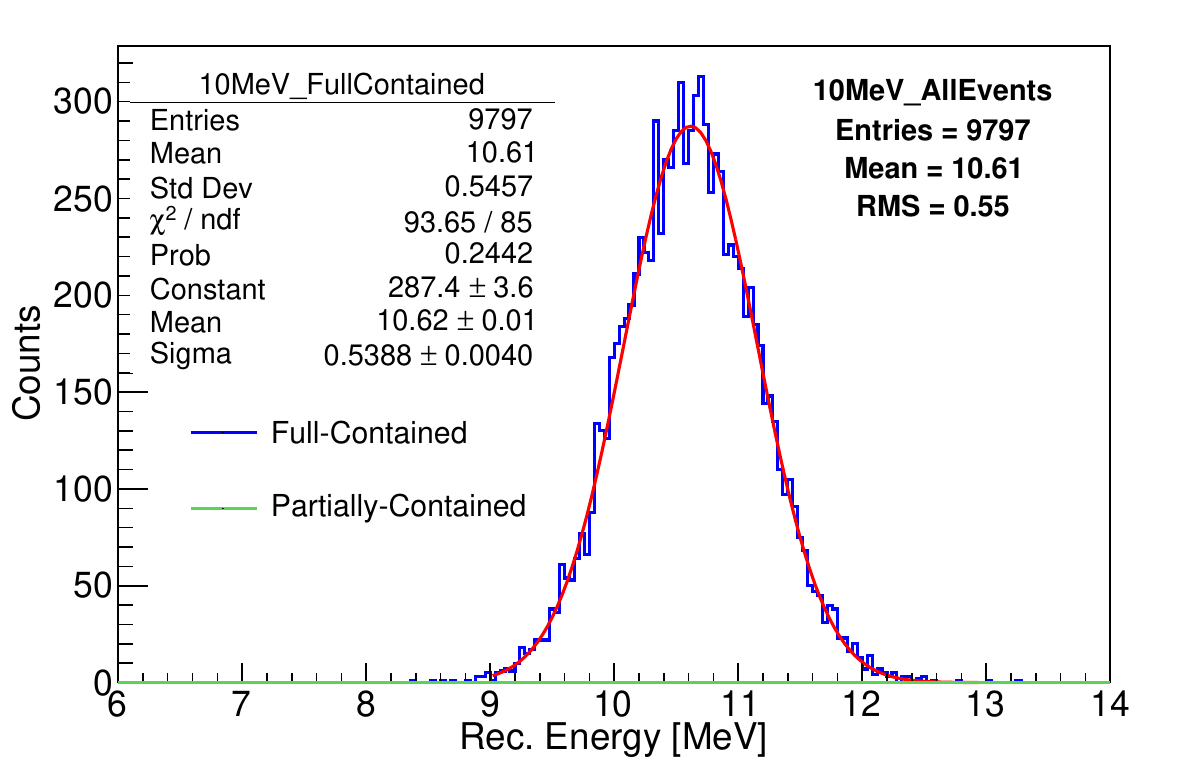}
    }
    \subfigure[50~MeV]
    {
    \label{Fig.sub.2 calib recv}
    \includegraphics[width=0.32\hsize]{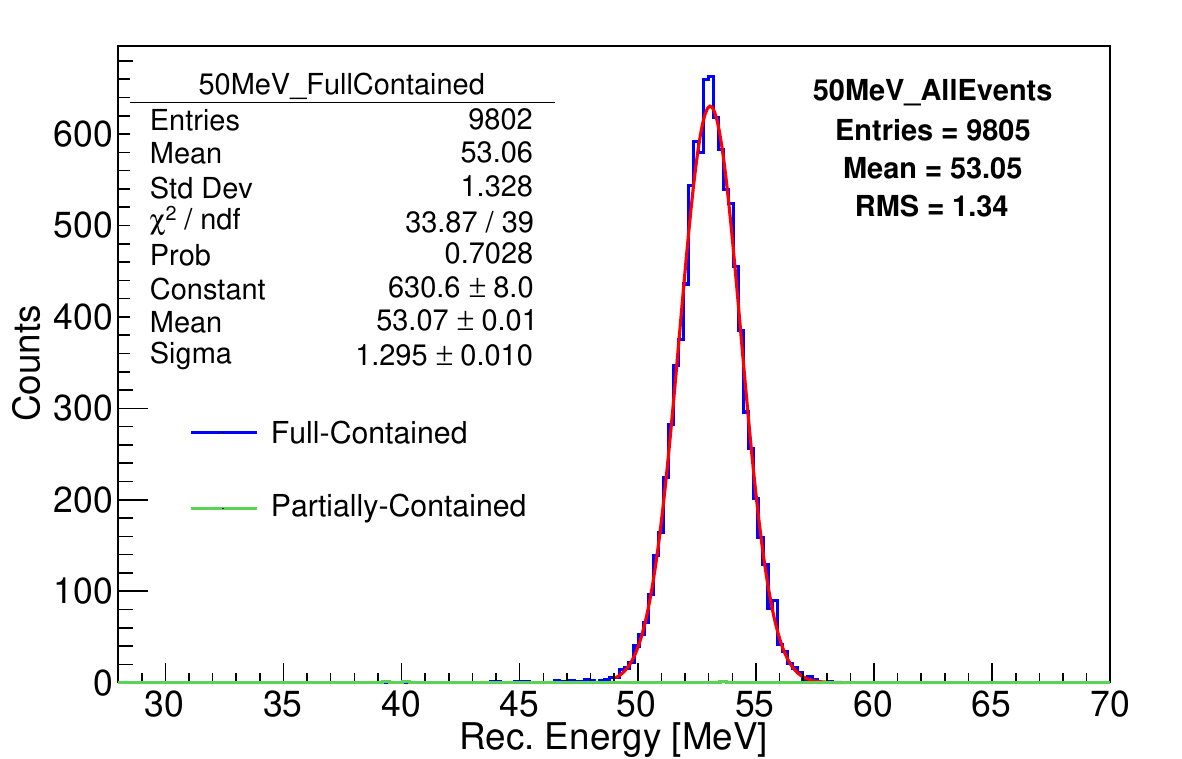}
    }
    \subfigure[200~MeV]
    {
    \label{Fig.sub.3 calib recv}
    \includegraphics[width=0.32\hsize]{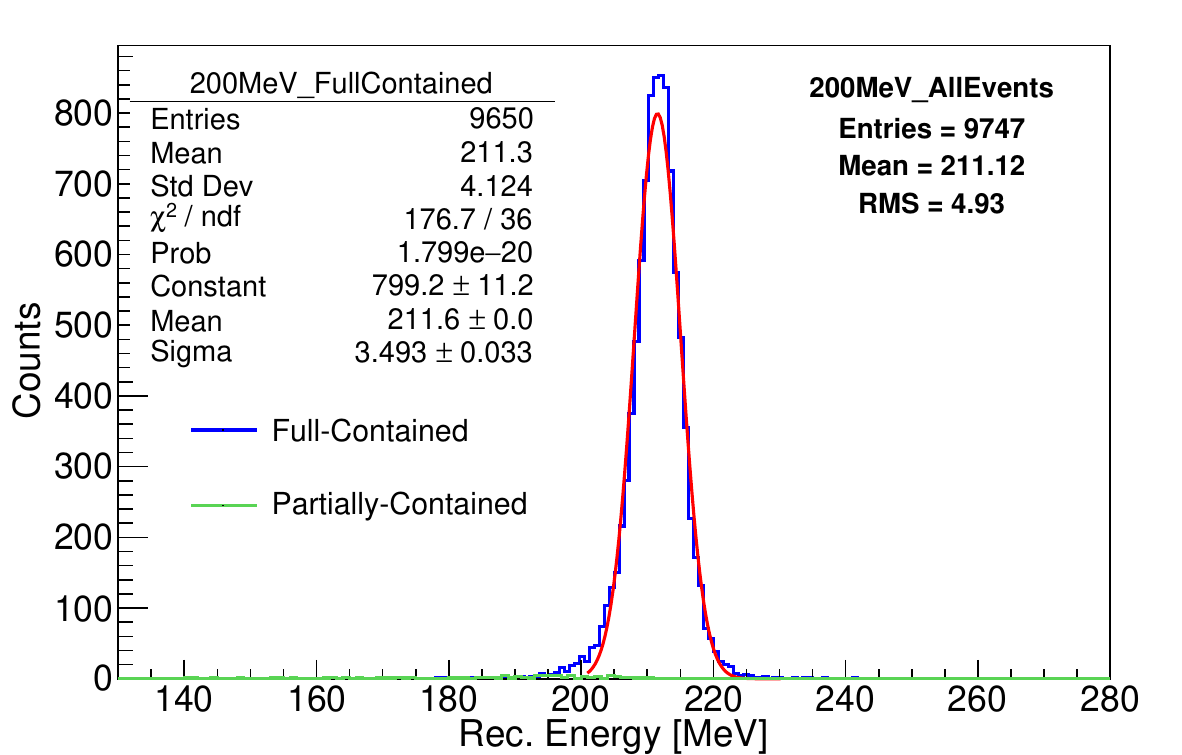}
    }
    \subfigure[500~MeV]
    {
    \label{Fig.sub.4 calib recv}
    \includegraphics[width=0.32\hsize]{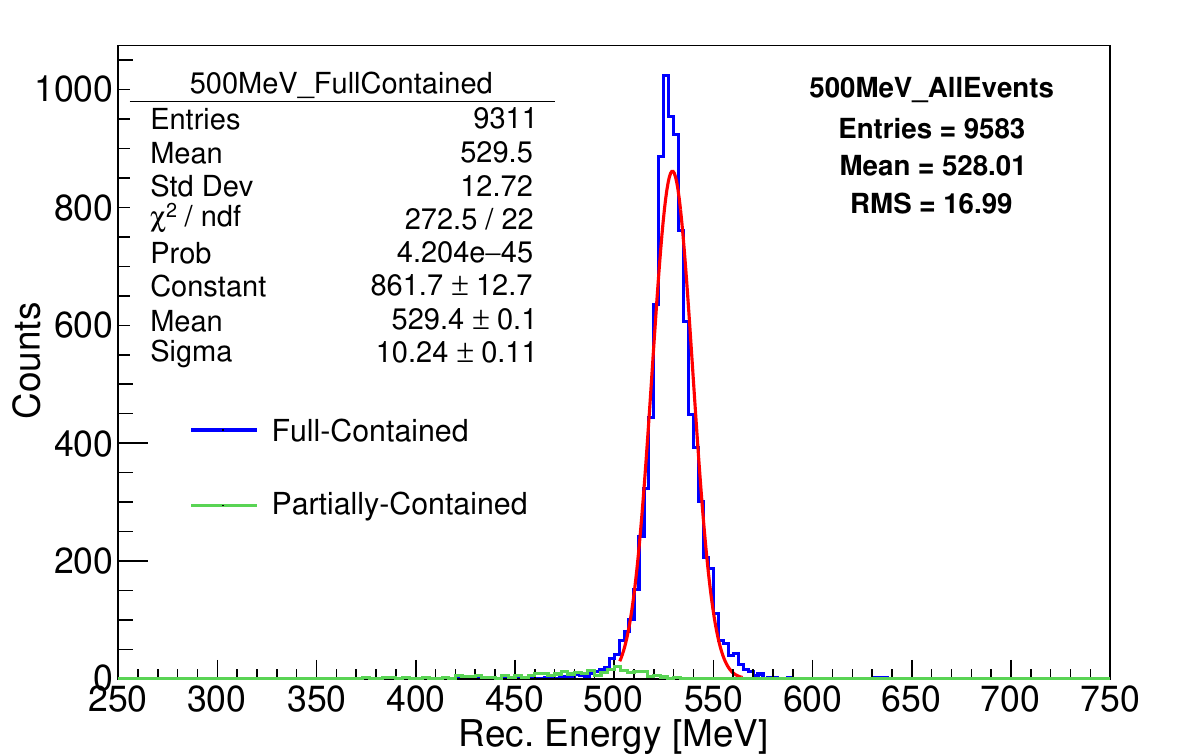}
    }
    \subfigure[700~MeV]
    {
    \label{Fig.sub.5 calib recv}
    \includegraphics[width=0.32\hsize]{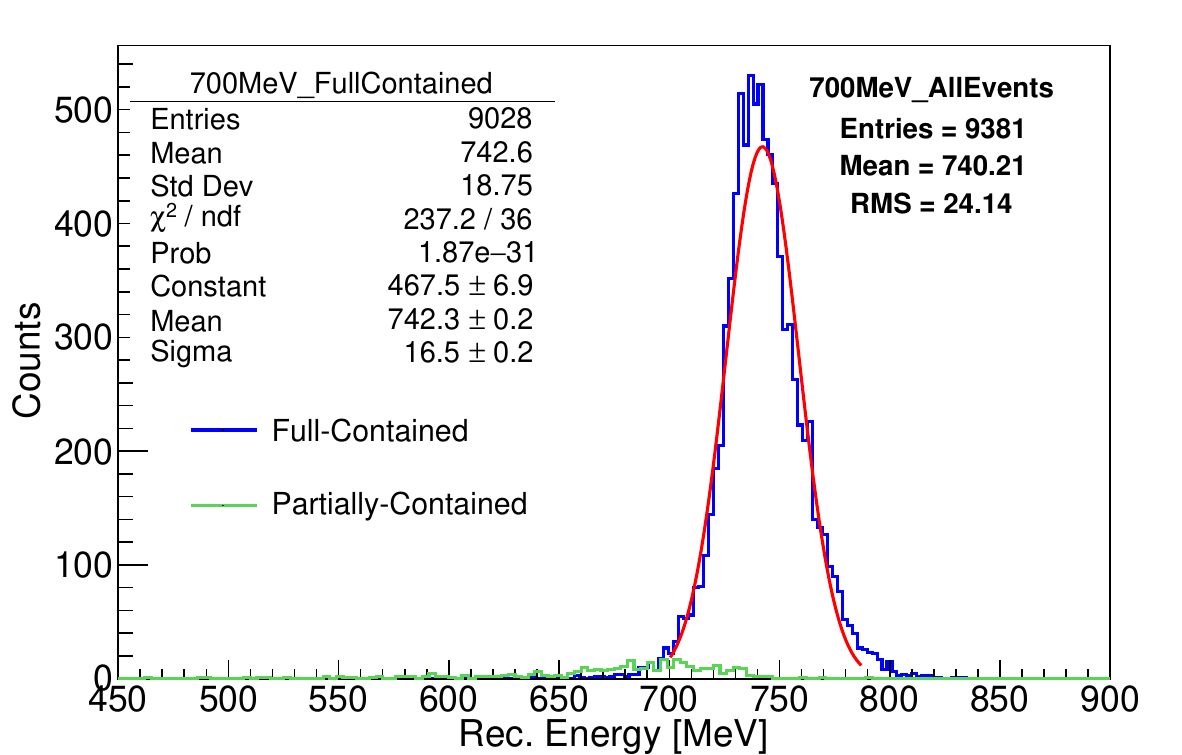}
    }
    \subfigure[1000~MeV]
    {
    \label{Fig.sub.6 calib recv}
    \includegraphics[width=0.32\hsize]{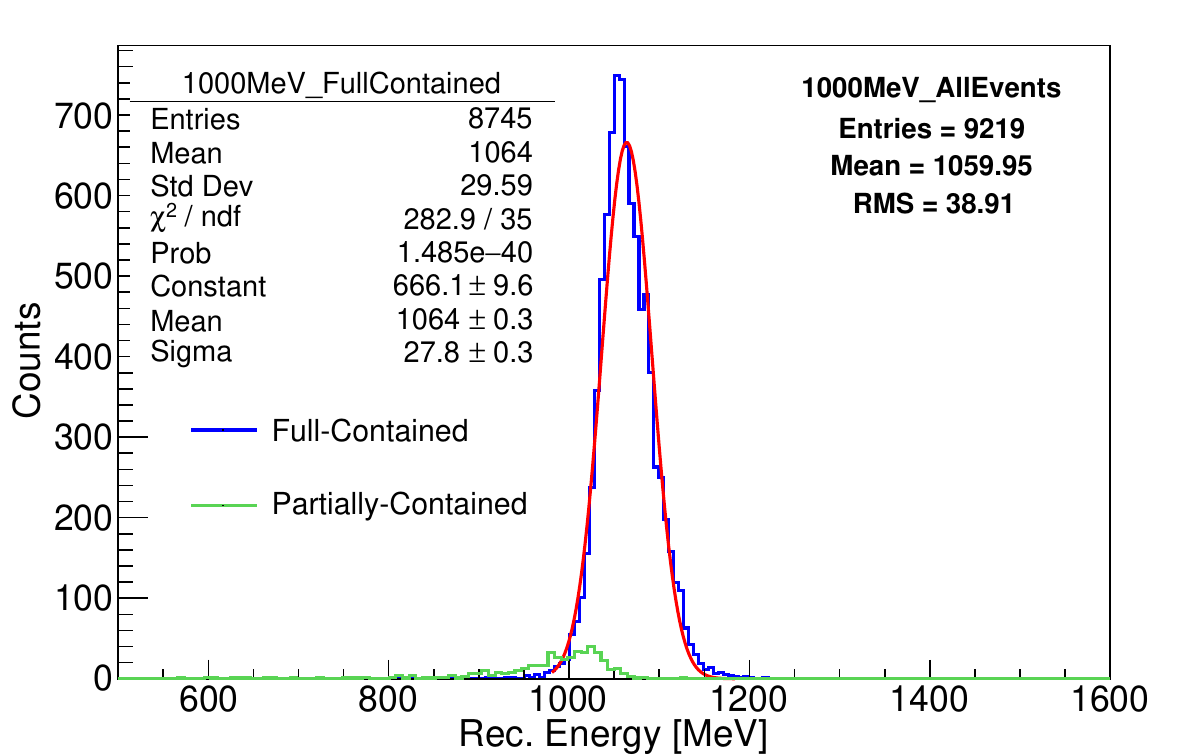}
    }
    \caption{Discrete energy reconstruction with reconstructed edep vertex after electronic simulation and charge reconstruction. The blue line is the FC events and the green line is the PC events. According to the fitting results (red line) of FC spectra, it can be observed that the reconstructed visible energy is about 6\% larger than the deposited energy of the electron. More specifically, for electrons with kinetic energies of 10~MeV, 50~MeV, 200~MeV, 500~MeV, 700~MeV and 1~GeV, the ratio of reconstructed visible energy to deposited energy is found to be 1.062, 1.061, 1.058, 1.059, 1.060 and 1.064, respectively. The corresponding explanation is provided in the text.}
    \label{Fig: calib with recv}
    \end{figure*}

\begin{figure*}[t]
    \centering  
    \subfigure[Without electronic simulation and charge reconstruction, using true energy-deposit center for energy reconstruction]
    {
    \label{Fig:detuniform_FCevent_sub1}
    \includegraphics[width=0.32\hsize]{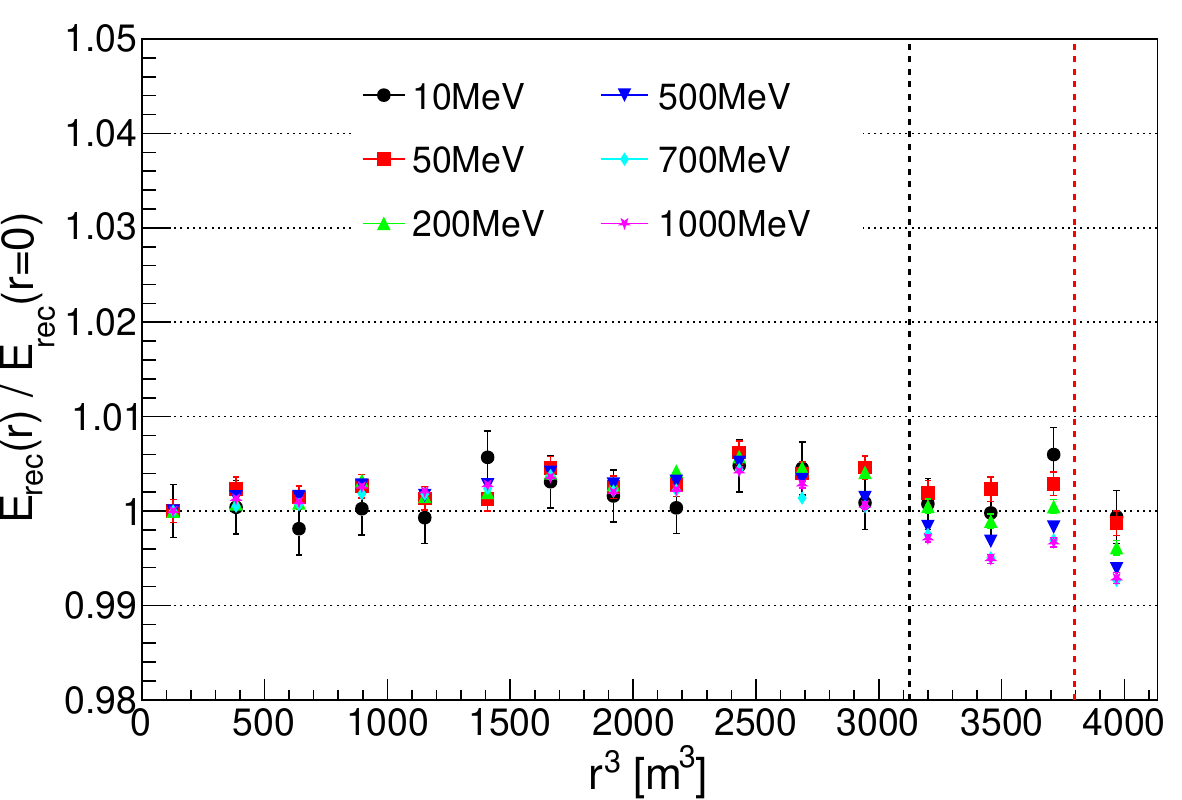}
    }
    \subfigure[With electronic simulation and charge reconstruction, using true energy-deposit center for energy reconstruction]
    {
    \label{Fig:detuniform_FCevent_sub2}
    \includegraphics[width=0.32\hsize]{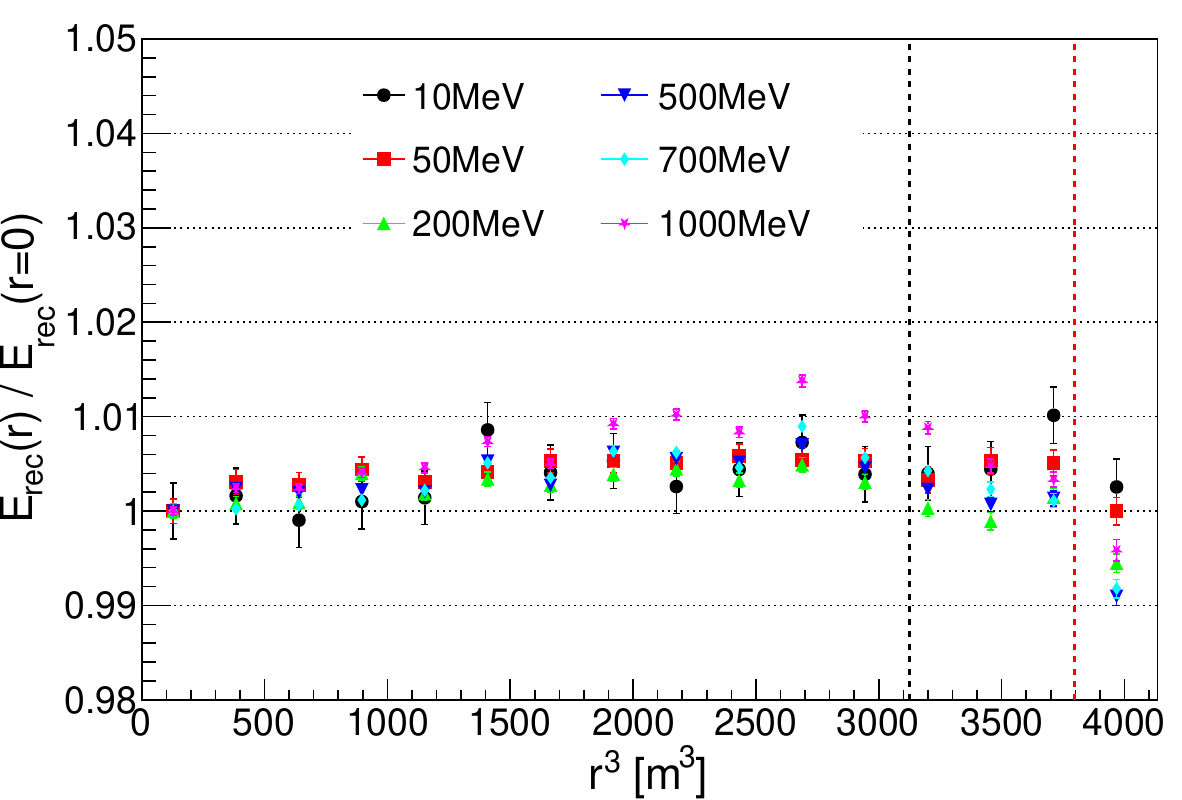}
    }
    \subfigure[With electronic simulation and charge reconstruction, using reconstructed energy-deposit center for energy reconstruction]
    {
    \label{Fig:detuniform_FCevent_sub3}
    \includegraphics[width=0.32\hsize]{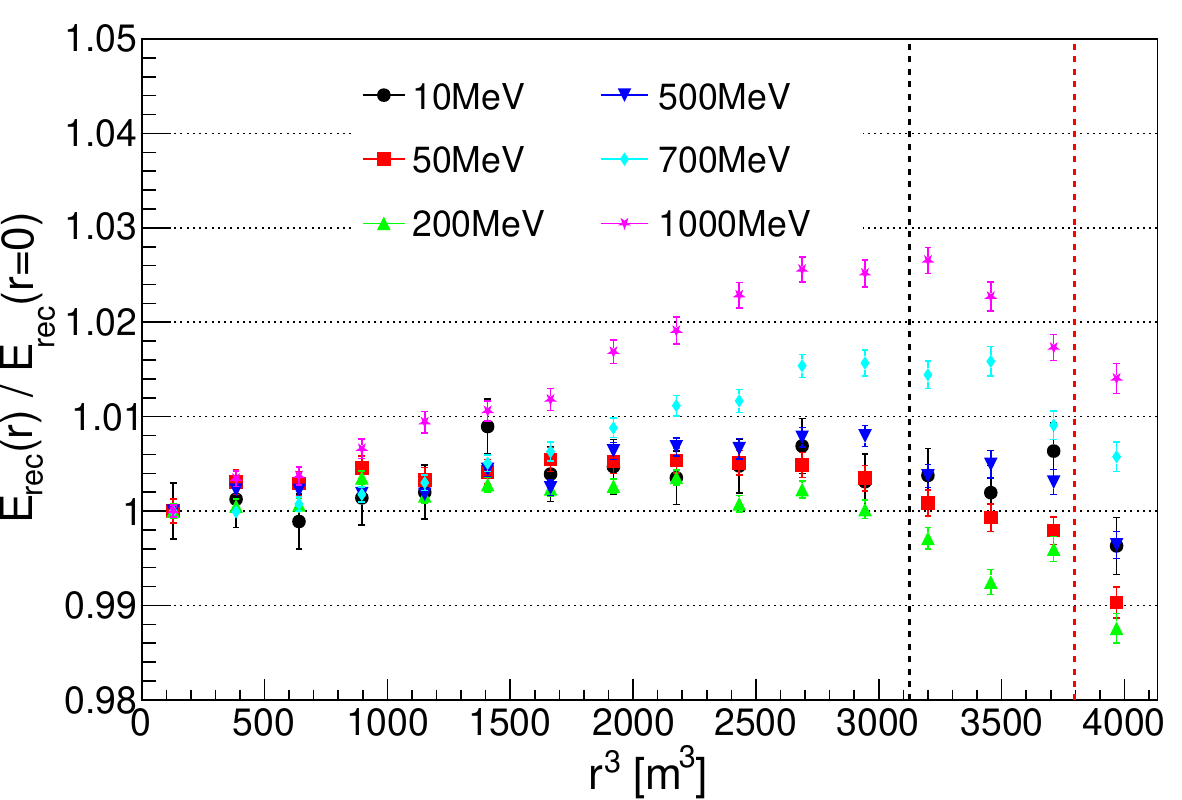}
    }
    \caption{Uniformity of discrete energy reconstruction for FC events. On each plot, black vertical dotted lines correspond to $r=15~m$, and red vertical dotted lines ($r=15.6$~m) correspond to the boundary of the total reflection region, which is caused by larger refractive index of the LS (which has a similar refractive index to the Acrylic) than water.}
    \label{Fig:detuniform_FCevent}
    \end{figure*}

    Next, the performance of energy reconstruction will be introduced. The reconstructed energy spectra for electrons with different kinetic energies are shown in Fig.~\ref{Fig: calib with recv}. The blue spectra correspond to events whose energies are fully contained (FC) in the LS, while the green spectra correspond to events whose energies are partially contained (PC) in the LS. For electrons with energies greater than 500~MeV, the proportion of PC events becomes larger and the 16~m cut can not totally exclude the case of energy leakage. The FC spectra can be well-fitted with a Gaussian function and the reconstructed energy is about 6\% larger than the deposited energy of the electron. According to the official simulation in JUNO, when anchored at the 2.223~MeV gamma peak generated by $(n,\gamma)$H, the high-energy electron has an energy non-linearity of $\sim$6\%~\cite{Yu-2022god}. Thus, this deviation is understood and it is mainly caused by the energy non-linearity response of LS. On the other hand, the non-uniformity of energy reconstruction may also introduce some small deviations, but generally less than 1\%, which is shown in Fig.\ref{Fig:detuniform_FCevent_sub3}. 

\begin{figure*}[t]
    \centering  
    \subfigure[Without electronic simulation and charge reconstruction, using true energy-deposit center for energy reconstruction]
    {
    \label{Fig:detuniform_allevent_sub1}
    \includegraphics[width=0.32\hsize]{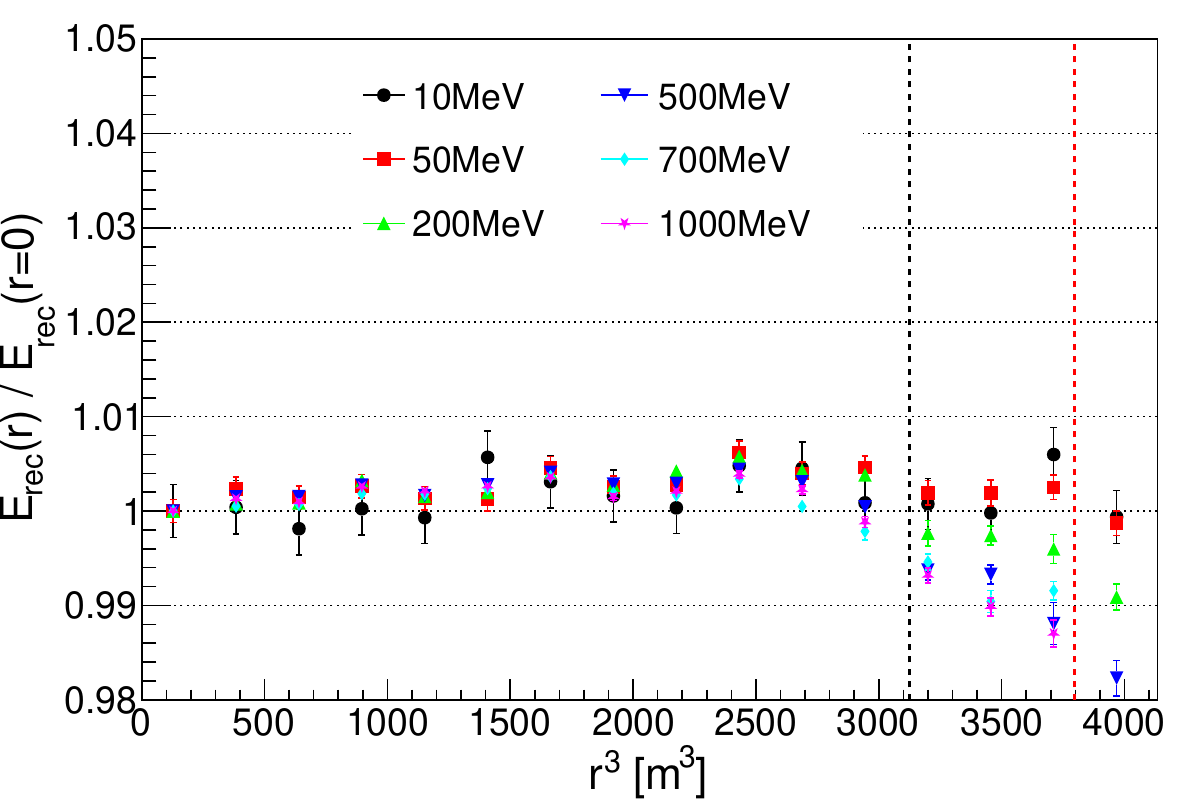}
    }
    \subfigure[With electronic simulation and charge reconstruction, using true energy-deposit center for energy reconstruction]
    {
    \label{Fig:detuniform_allevent_sub2}
    \includegraphics[width=0.32\hsize]{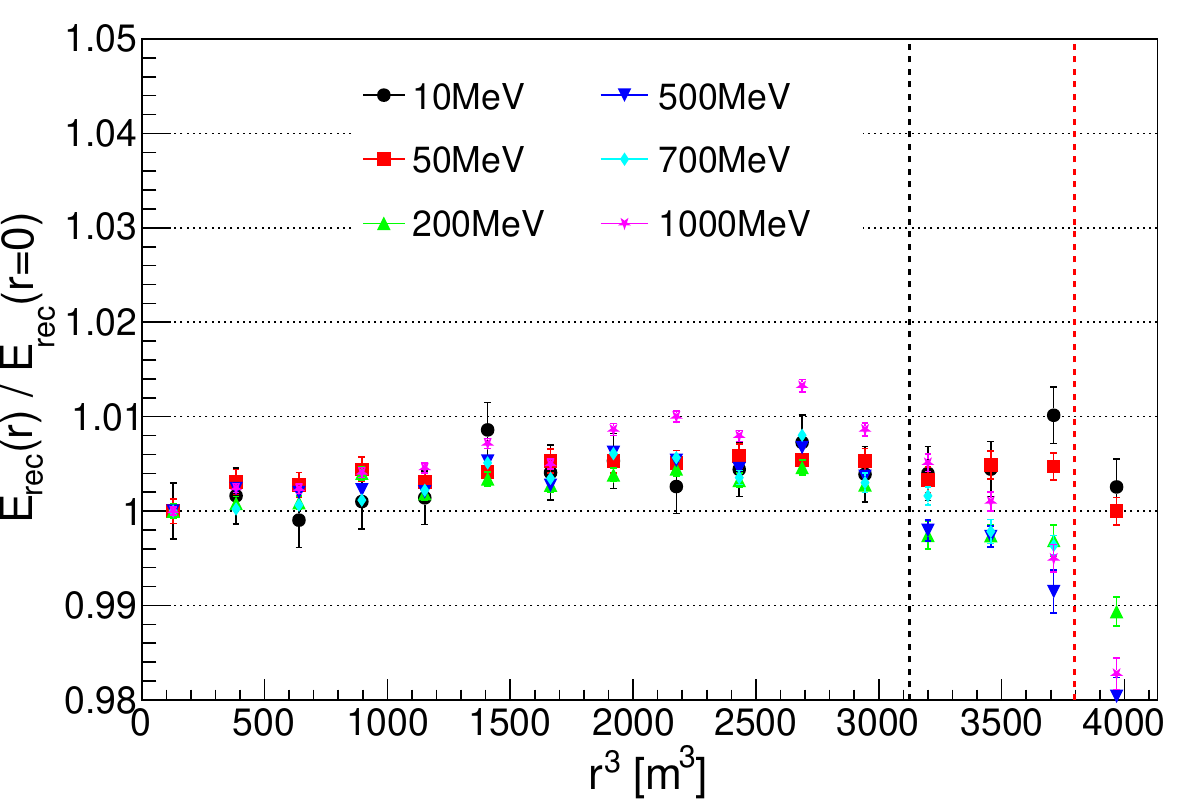}
    }
    \subfigure[With electronic simulation and charge reconstruction, using reconstructed energy-deposit center for energy reconstruction]
    {
    \label{Fig:detuniform_allevent_sub3}
    \includegraphics[width=0.32\hsize]{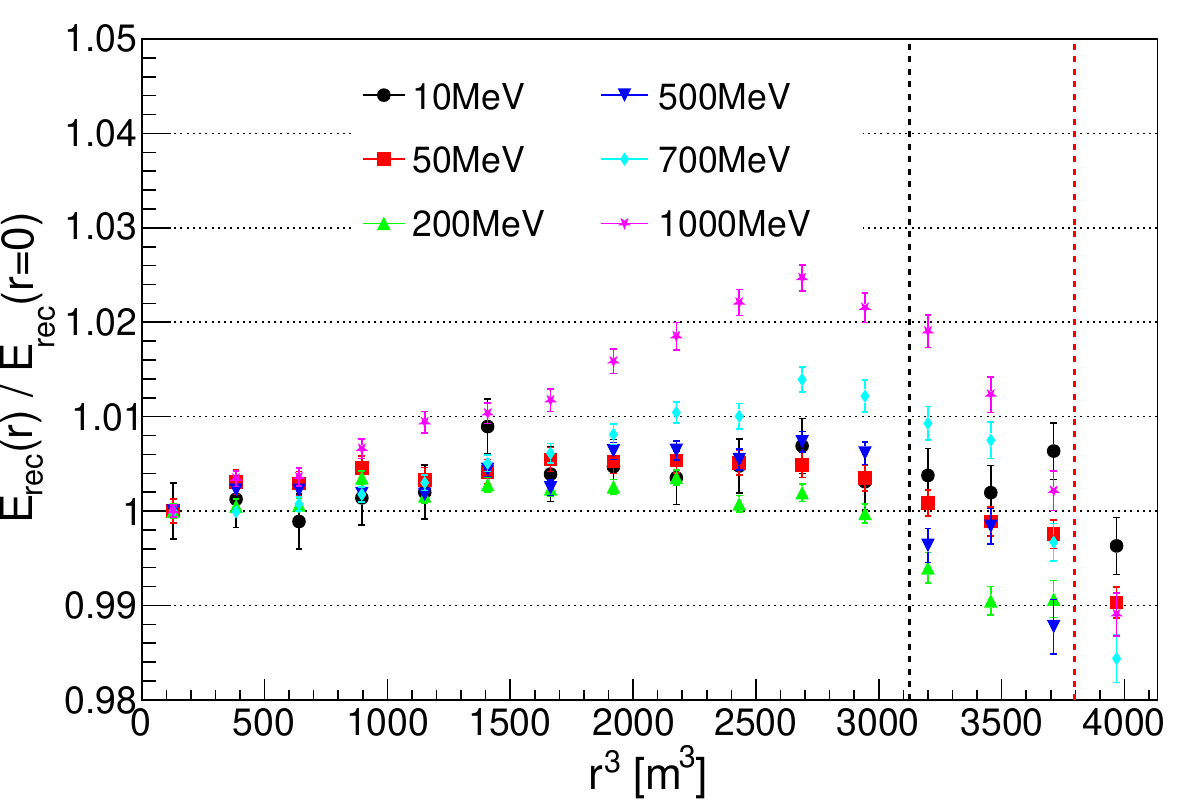}
    }
    \caption{Uniformity of discrete energy reconstruction for all (FC+PC) events. On each plot, black vertical dotted lines correspond to $r=15~m$, and red vertical dotted lines ($r=15.6$~m) correspond to the boundary of the total reflection region, which is caused by larger refractive index of the LS (which has a similar refractive index to the Acrylic) than water.}
    \label{Fig:detuniform_allevent}
    \end{figure*}

    To understand the non-uniformity shown in Fig.\ref{Fig:detuniform_FCevent_sub3}, Fig.\ref{Fig:detuniform_FCevent_sub1} and Fig.\ref{Fig:detuniform_FCevent_sub2} can be compared, which corresponds to the cases using true energy-deposit center without/with electronic simulation and charge reconstruction. In Fig.\ref{Fig:detuniform_FCevent_sub1}, for electron samples with different energies, the non-uniformity is consistent at about 0.5\% from the center of the detector to the edge. After electronics simulation and charge reconstruction (Fig.\ref{Fig:detuniform_FCevent_sub2}), there is a slight increase in non-uniformity, but it still remains within 1.5\%. Fig.\ref{Fig:detuniform_FCevent_sub3} corresponds to the case using reconstructed energy-deposit center which shows deviation (Fig.\ref{Fig:vertexrec}), as a result, for 500~MeV and 1~GeV electrons, their energy non-uniform are about 2\% and 3\% at the edge. Furthermore, if the PC events are included (Fig.~\ref{Fig:detuniform_allevent}), they mainly affect the non-uniformity of high-energy electrons located in the edge region.


    \begin{figure}[!htb]
        \centering  
        \includegraphics[width=0.9\hsize]{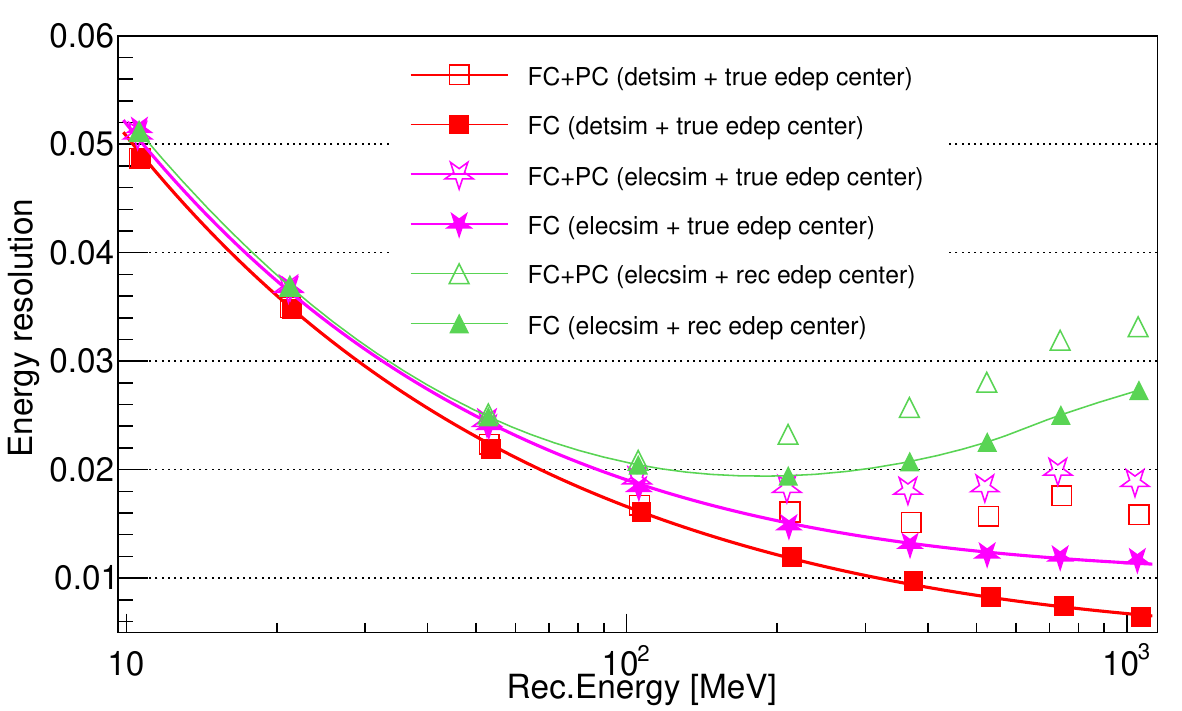}
        \caption{Resolution in several discrete energy in simulation phase. The red squares correspond to the resolution in the detector simulation. The pink stars and the green triangles correspond to the resolution after electronic simulation and charge reconstruction. The energy reconstructions of the red squares and pink stars use the true center of energy deposition, while the green triangles utilize the reconstructed center for their energy reconstruction. The energy reconstruction of the red squares and the pink stars uses the true energy-deposit center, while the energy reconstruction of the green triangles uses the reconstructed energy-deposit center. On the other hand, solid and hollow are used to mark the FC events and FC+PC events, respectively}
        \label{res: detresolution}
    \end{figure}

\begin{figure}[!htb]
        \centering 
        \includegraphics[width=0.9\hsize]{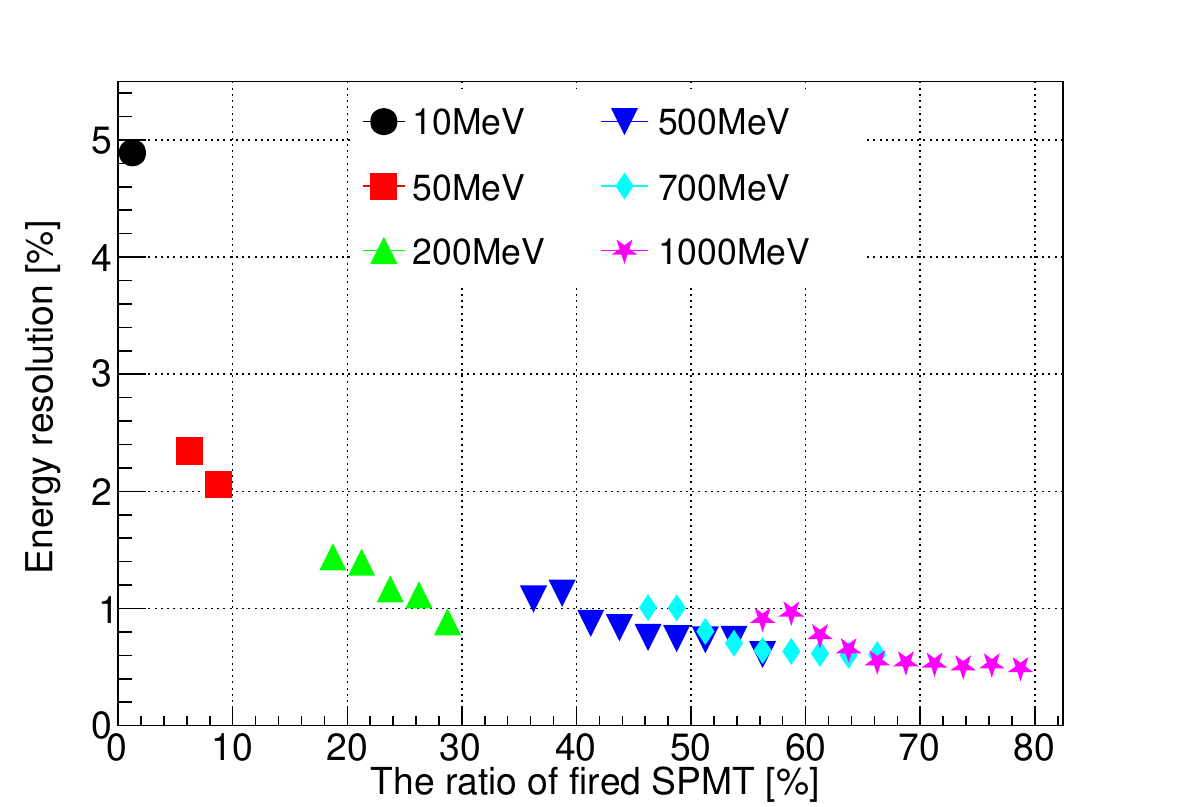}
        \caption{The resolution varies with the fired ratio of SPMT at different energies}
        \label{Fig:firedRes}
    \end{figure}

    Figure~\ref{res: detresolution} shows the performance of energy resolution. The solid points correspond to the reconstruction results of FC events, while the hollow points include both FC events and PC events. The red squares and pink stars denote the cases using the true energy-deposit center for energy reconstruction, while the green triangles use the reconstructed energy-deposit center. In addition, electronics simulation and charge reconstruction are applied in the reconstruction result shown by the pink stars and the green triangles. Comparing energy resolutions in different conditions, it can be found that the energy reconstruction performance of the high-energy events is good using the OCCUPANCY strategy and the energy resolution is about 0.8\% for 1~GeV electrons in the ideal case (red solid squares). For a more realistic situation by including electronic simulation and charge reconstruction, the resolution is only about 0.3\% worse which indicates that the correction works well in controlling the influence of PMT dark count and threshold effect. From the comparison of solid points and hollow points, the PC events mainly affect the electrons whose kinetic energy is larger than 100~MeV and their energy resolutions will deteriorate by about 1\%. In real detection, the reconstructed energy-deposit center is required for energy reconstruction, and its smearing will introduce additional smearing on the reconstructed energy, especially for the high-energy electrons. As a result, the energy resolution is about 3.2\% for 1~GeV electrons based on our algorithm. In Fig.~\ref{Fig:firedRes}, the relationship between energy resolution and the fired ratio of SPMT is investigated using the electron sample without electronics simulation, and true energy-deposit center is applied. In general, a higher fired ratio of SPMT corresponds to a better energy resolution. This indicates that our algorithm has the potential to be applied to higher energy events when the fired ratio of SPMT is not close to 1, but must solve the problem of energy-deposit center reconstruction which has a larger bias at higher energies.
    
\section{Summary}
\label{section:Summary}
    Accurate energy reconstruction is crucial to detect various physics events in a wide energy range from MeV to GeV in JUNO. This work focuses on the energy reconstruction of sub-GeV events based on 3-inch PMTs and the OCCUPANCY method. Our reconstruction shows good performances in the test of MC simulation samples. The energy non-uniformity can be controlled within 1\% from the center of the detector to the edge for electrons whose kinetic energies are smaller than 500~MeV. As for 1~GeV electron, the energy non-uniformity can be controlled within 3\%. The energy resolutions for 1~GeV electron FC events and FC+PC events are about 2.7\% and 3.2\%, respectively. Our algorithm has the advantage of small dependency on precise charge reconstruction by mainly using the information from SPMT’s firing states.


 \end{document}